%% file: acs-main.tex
\documentclass[journal=jcisd8, manuscript=article, layout=twocolumn]{achemso}

\pdfoutput=1 
\usepackage[version=3]{mhchem} 
\usepackage{multirow}
\usepackage{graphicx}
\graphicspath{{images/}}      
\usepackage{hyperref}
\hypersetup{colorlinks=true,urlcolor=blue,citecolor=blue,linkcolor=black}

\newcommand{\repourl}[0]{\url{https://github.com/ulissigroup/wherewulff}}
\newcommand{\angstrom}{\mbox{\normalfont\AA}}
\newcommand{\dataurl}[0]{\url{https://doi.org/10.5281/zenodo.7600476}}

\author{Rohan Yuri Sanspeur}
\affiliation{Department of Chemical Engineering, Carnegie Mellon University}
\altaffiliation{These authors contributed equally to this work}
\author{Javier Heras-Domingo}
\affiliation{Department of Chemical Engineering, Carnegie Mellon University}
\altaffiliation{These authors contributed equally to this work}
\author{John R. Kitchin}
\affiliation{Department of Chemical Engineering, Carnegie Mellon University}
\author{Zachary Ulissi}
\affiliation{Department of Chemical Engineering, Carnegie Mellon University}
\alsoaffiliation{Scott Institute for Energy Innovation, Carnegie Mellon University}
\email{zulissi@andrew.cmu.edu}

\title[]
    {WhereWulff: A semi-autonomous workflow for systematic catalyst surface reactivity under reaction conditions}

\abbreviations{IR,NMR,UV}

\let\oldmaketitle\maketitle
\let\maketitle\relax

\begin{document}

\begin{tocentry}
\includegraphics{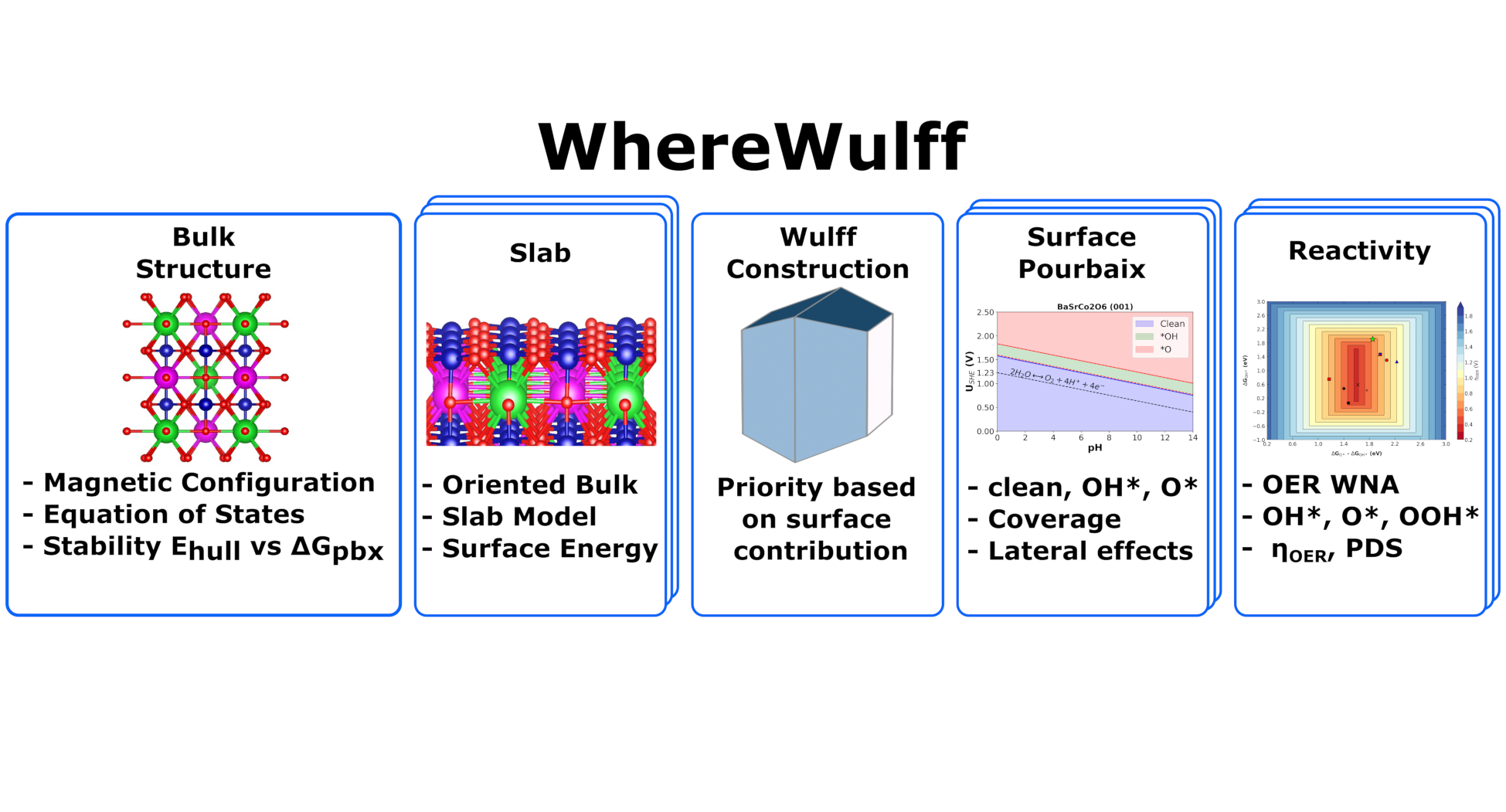}
\end{tocentry}

\twocolumn[
\begin{@twocolumnfalse}
\oldmaketitle
\input{sections/abstract}
\end{@twocolumnfalse}
]

\clearpage

\input{sections/intro}
\input{sections/methods}
\input{sections/results}

\input{sections/conclusions}

\begin{acknowledgement}
The authors thank the National Research Council of Canada (NRC) project Award MCF-111 and the Army Research Office (ARO) project Award W911NF2010188. The authors thank Richard Tran for fruitful discussions and feedback as we developed this framework. This research also used resources at the National Energy Research Scientific Computing Center (NERSC), a U.S. Department of Energy Office of Science User Facility located at Lawrence Berkeley National Laboratory.
\end{acknowledgement}

\begin{suppinfo}
The supporting information contains details on the general DFT method used in this work along with the surface energy derivation for non-stoichiometric surfaces, schematics of the auxiliary bulk structure workflow, DFT figures of the associated Pourbaix and OER intermediates for all the systems explored in this work, surface Pourbaix diagrams and workflow deployment strategies. In the vein of reproducibility, we have made available the electronic structure data and metadata at \dataurl.
\end{suppinfo}

\bibliography{acs-references.bib, esi-reference.bib}

\end{document}


\clearpage
\small
\tableofcontents
\clearpage


\section{Non-stoichiometric Surface Energy Calculations}
One can compute the surface energy of a symmetric slab:
\begin{equation}
    \gamma = \frac{\bigl[G_{\text{slab}} - \sum_{i}N_i\mu_i\bigr]}{2A}
    \label{surface_energy_symmetric}
\end{equation}
distributing it between the two surfaces, each having area $A$, where the sum is over the chemical potentials $\mu_i$ and the number of atoms $N_i$ of each of the species that make up the composition of the slab, as shown in Eq. (\ref{surface_energy_symmetric}).

Assuming that the bulk is in equilibrium with the slab and by defining a reference, we can conveniently re-write Eq. (\ref{surface_energy_symmetric}) to be as a function of the bulk energy per formula unit, as shown in Eq. (\ref{bulk_energy_formula_unit}), where the last sum represents the free energy excess, $x_{i}$ the number of atoms per bulk formula and $N_\text{ref}$ is the reference specie that is picked.
\begin{equation}
    \gamma = \frac{\bigl[G_{\text{slab}} - N_{\text{ref}} \cdot g_{\text{bulk}} - \sum_{i}\bigl(N_i - x_{i} \cdot N_{\text{{ref}}}\bigr) \cdot \mu_{i}\bigr]}{2A}
    \label{bulk_energy_formula_unit}
\end{equation}

More concretely, the derivation of Eq.~(\ref{bulk_energy_formula_unit}) for a slab consisting of $N_{Ba}$, $N_{Ti}$, $N_{O}$ and a bulk with formula $\ce{BaTiO4}$:
\begin{itemize}

    \item From first principles we can express the Gibbs energy of the system as the sum of the interfacial energy and the cost of introducing/removing elemental species$_i$ at some temperature (T) and pressure (p)
    \begin{equation}
        \begin{aligned}
        G(x_{Ba}, x_{Ti}, x_{O}, T, p) = \mu_{Ba}(T,p)N_{Ba} + \\ \mu_{Ti}(T,p)N_{Ti} + \\ \mu_{O}(T,p)N_{O} + \\ \gamma \cdot 2A
        \end{aligned}
        \label{eq1}
    \end{equation}
  \item Rearranging Eq.~(\ref{eq1}) for the surface energy, $\gamma$, and assuming symmetric surfaces, we get Eq.~(\ref{eq2})
    \begin{equation}
        \begin{aligned}
        \gamma = \frac{
        \begin{matrix}
        \bigl[G(x_{Ba}, x_{Ti}, x_{O}, T, p) - \mu_{Ba}(T,p)N_{Ba} - &\\ \mu_{Ti}(T,p)N_{Ti} - \mu_{O}(T,p)N_{O}\bigr]
        \end{matrix}}{2A}
        \end{aligned}
        \label{eq2}
    \end{equation}
    \item If we assume that the bulk is in equilibrium with the slab, we can write Eq.~(\ref{eq3})
    \begin{equation}
        \begin{aligned}
        g_{\ce{BaTiO4}} = \left[x_{Ba}^{bulk}\mu_{Ba}(T,p) + x_{Ti}^{bulk}\mu_{Ti}(T,p) + x_{O}^{bulk}\mu_{O}(T,p)\right]\cdot N_{bulk}
        \end{aligned}
    \label{eq3}
    \end{equation}
    \item We define the reference to be in relation to Ba per Eq.~(\ref{eq4})
    \begin{equation}
        \begin{aligned}
        \mu_{Ba} = \frac{1}{x_{Ba}^{bulk}}\left(\frac{g_{\ce{BaTiO4}}}{N_{bulk}} - x_{O}^{bulk}\mu_{O}(T,p) - x_{Ti}^{bulk}\mu_{Ti}(T,p)\right)
        \end{aligned}
        \label{eq4}
    \end{equation}
    \item We then introduce the bulk energy per formula unit into Eq.~(\ref{eq2}) by substituting Eq.~(\ref{eq4}), to get Eq.~(\ref{eq5})
    \begin{equation}
        \begin{aligned}
        \begin{matrix}
        \gamma = \frac{1}{2A}\bigl[G(x_{Ba}, x_{Ti}, x_{O}, T, p) - \frac{N_{Ba}}{x_{Ba}^{bulk}N_{bulk}}g_{\ce{BaTiO4}} + & \\ \mu_{O}(T,p)\left(\frac{x_{O}^{bulk}N_{Ba}}{x_{Ba}^{bulk}} - N_{O}\right) + \mu_{Ti}(T,p)\left(\frac{x_{Ti}^{bulk}N_{Ba}}{x_{Ba}^{bulk}} - N_{Ti}\right)\bigr]
        \end{matrix}
        \end{aligned}
    \label{eq5}
    \end{equation}
    \item Eq.~(\ref{eq5}) reduces to the well-known surface energy Eq.~(\ref{eq6}) when the stoichiometry between the bulk and the interface is maintained
    \begin{equation}
        \begin{aligned}
        \gamma = \frac{E_{slab} - n \cdot E_{bulk}}{2A}
        \end{aligned}
        \label{eq6}
    \end{equation}
\end{itemize}
    
The slab model is made up of a 2D surface and a corresponding oriented bulk, since this has been shown to most efficiently converge the surface energy calculations~\cite{sun_ceder_2013}.
\section{Surface Coverage and OER}
The coverage effects of reaction intermediates (OH*, O*) may significantly impact the local environment of the active site, resulting in changes in the adsorption strength of the OER reaction intermediates. Consequently, it is important to determine the most stable surface coverage at given conditions of applied potential and pH to more reasonably describe the OER activity. To determine the most stable surface coverage, Surface Pourbaix diagrams\cite{venkat}, were constructed at three extreme coverages, clean, \ce{OH^{*}}, and \ce{O^{*}} terminated.

OER catalytic activities of the different surface structures were determined by the theoretical overpotential ($\eta_{OER}$) and the potential-determining step (PDS) by assuming an associative reaction mechanism with O*, OH* and OOH* as reaction intermediates. The four proton-coupled electron transfers (PCET) reactions under acidic conditions are:

\begin{equation}
    \label{eq:oer_1}
    \ce{H2O + ($\ast$) -> OH^* + (H^+ + e^-)}
\end{equation}
\begin{equation}
    \ce{OH^* -> O^* + (H^+ + e^-)}
\end{equation}
\begin{equation}
    \ce{O^* + H2O -> OOH^* (H^+ + e^-)}
\end{equation}
\begin{equation}
    \label{eq:oer_end}
    \ce{OOH^* -> O2 (^) + ($\ast$) + (H^+ + e^-)}
\end{equation}

The computational hydrogen electrode (CHE) was used to express the chemical potential of the proton-electron pair (\ce{H^+ + e^-}), which is related to the chemical potential of \ce{H2} based on the equilibrium $\mu[\ce{H^+}]+\mu[\ce{e^-}]= 0.5\; \mu[\ce{H2}(g)]$ at 0 V$_{RHE}$ (Reversible Hydrogen Electrode) and corrects the driving force with the deviation of the applied potential from the equilibrium situation. To avoid the use of \ce{O2} electronic energy, which is difficult to determine correctly within standard GGA-DFT, the experimental free energy of \ce{2H2O -> O2 + 2H2} ($\Delta G = 4.92$ eV), was used. Therefore, the Gibbs free energies of reactions (\ref{eq:oer_1})-(\ref{eq:oer_end}) depend on the adsorption free energies of the reaction intermediates ($\Delta G_{OH*}$, $\Delta G_{O*}$, and $\Delta G_{OOH*}$) which are calculated relative to \ce{H2O}(g) and \ce{H2}(g) at $U = 0$ V and standard conditions. In total, 133 adsorption calculations were performed to calculate the adsorption free energies of the reaction intermediates, considering multiple orientations for \ce{OH^{*}} and \ce{OOH^{*}} on the surface termination for all the materials, and selecting the most stable one in each case.

The theoretical thermodynamic OER overpotential, which is a measure of the activity of a catalyst, is then defined from Gibbs free energies of reactions (\ref{eq:oer_1})-(\ref{eq:oer_end}):
\begin{equation}
    \label{eq:overpotential}
    \eta_{OER}(V) = max[\Delta G_{\ce{OH}}, \Delta G_{\ce{O}} - \Delta G_{\ce{OH}}, \Delta G_{\ce{OOH}} - \Delta G_{\ce{O}}, 4.92 - \Delta G_{\ce{OOH}}] /e - 1.23
\end{equation}
The step with the largest value in Eq. (\ref{eq:overpotential}) is referred to as the potential-determining step (PDS). It is important to note that the overpotential should not be compared directly with a measured overpotential, since the measured overpotential depends on the current density.

\section{Convergence Tests and Settings}
\begin{figure}[H]
    \centering
    \includegraphics[width=1\textwidth]{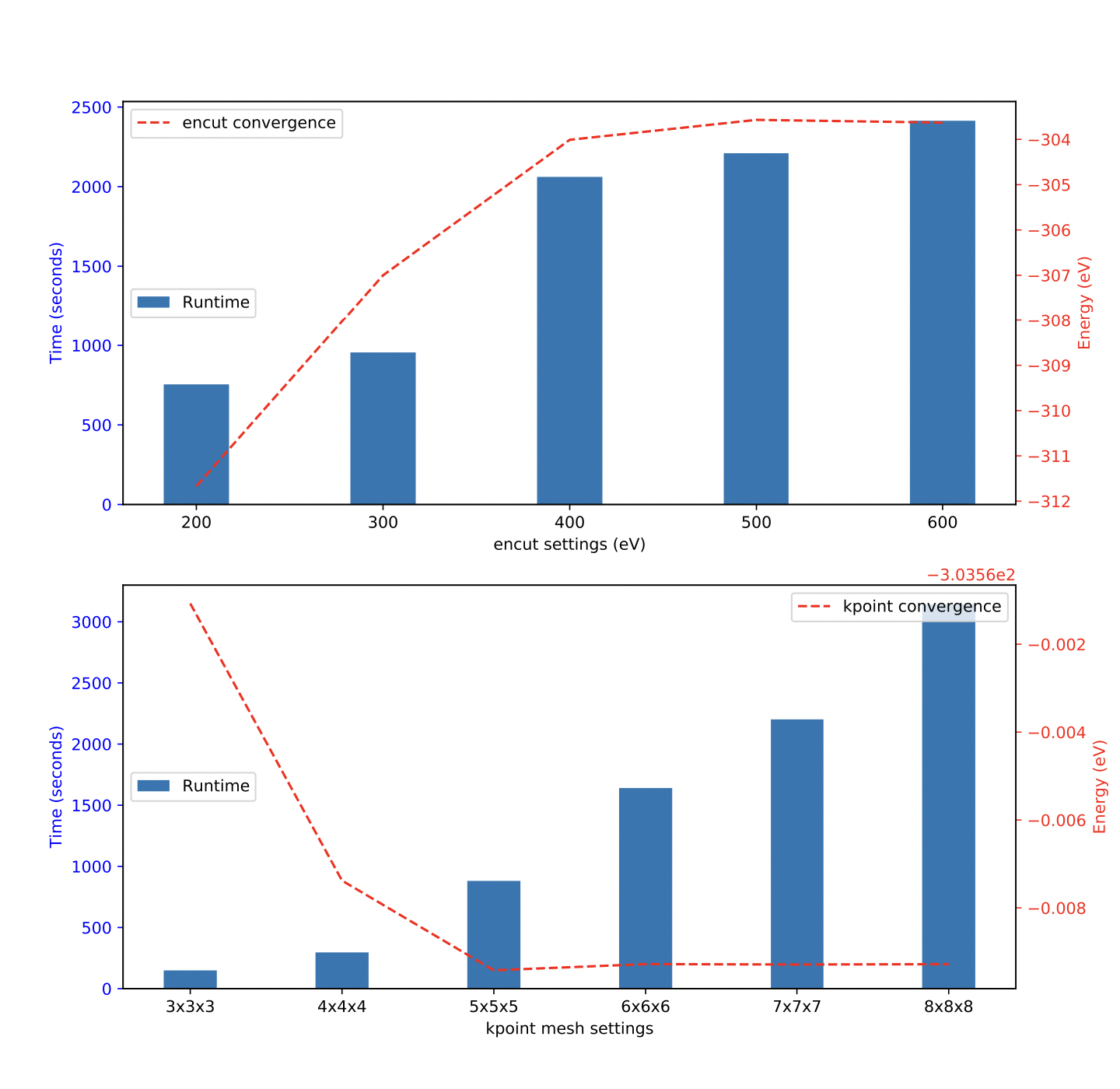}
    \caption{Standard convergence tests on bulk \ce{BaSnTi2O6} illustrating the trade-off between accuracy and computational time. Based on these results, we infer that using an ENCUT setting of 500 eV and a kpoint grid of 5$\times$5$\times$5 would be a good balance between accuracy and runtime. Our custom DFT settings, implemented as a child class of Pymatgen's \textit{MVLSlabSet}, obey these thresholds. }
    \label{fig:my_label}
\end{figure}

\begin{table}[]
    \centering
    \begin{minipage}{\textwidth}
    \begin{tabular}{c|c|c|c|c}
         \hline
         Slab Thickness & Slab Energy & $\gamma_{(hkl)}$ & Runtime & Number of atoms\\
         (multiple of oriented unit cell) & (eV) & $\left(\frac{J}{m^2}\right)$ & (days) & \\
         \hline\hline
         1 & -308.70 & NA & 0.12 & 40\\
         2 & -610.81 & 0.60 & 3.79 & 80\\
         3 & -919.93 & 0.56 & 15.73 & 120 \\
         4 & -1228.37 & 0.58 & 16.89 & 160 \\
         5\footnote{This slab optimization did not converge in the allotted number of re-submissions (10). These results corroborate that it is infeasible for us to carry out the slab optimizations at those levels of thickness and under such large numbers of degrees of freedom, which are shown to trigger multiple re-submissions (Figure~\ref{Slab_thick_webgui}).} & -1536.79 & 0.61 & 15.79 & 200\\
         \hline
    \end{tabular}
    \end{minipage}
    \caption{Slab thickness convergence table showing how we settled on a slab thickness of 3 in units of oriented unit cell under a (2 $\times$ 1) supercell. Since the surface energy changes within 1.24 $\frac{meV}{\angstrom^2}$ from a slab thickness of 3 to 4 and the runtime cubically in the number of electrons, we compromise on a slab thickness of 3 consistently throughout our case study.}
    \label{tab:my_label}
\end{table}

\begin{figure}
    \centering
    \includegraphics[width=\textwidth]{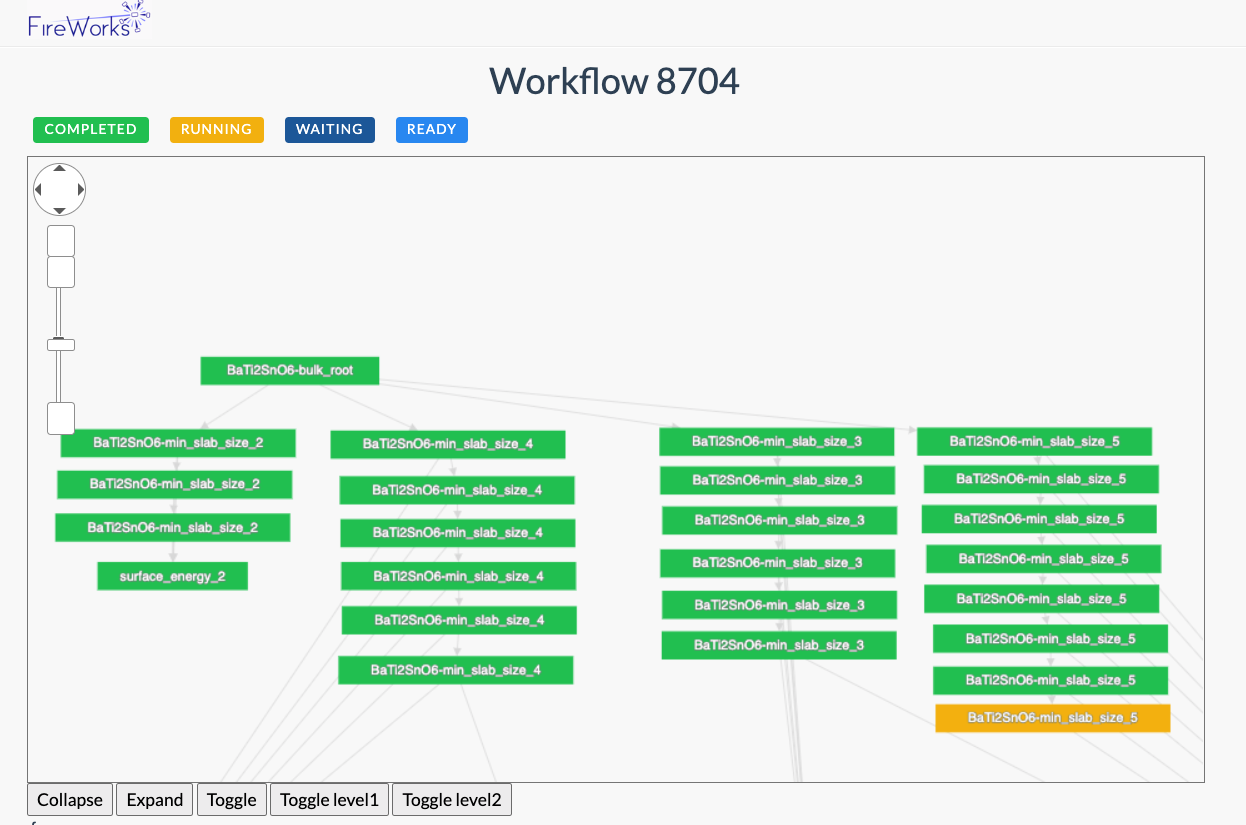}
    \caption{Fireworks~\cite{fireworks} web-gui showing how we repurposed some of the components of \textit{WhereWulff} to conduct slab thickness convergence tests.}
    \label{Slab_thick_webgui}
\end{figure}
\section{De-prioritized Surface Energies}
\begin{table}[H]
    \centering
    \begin{tabular}{c|c|c}
         \hline
         Formula & (hkl) & $\gamma_{(hkl)}$ \\
         \hline \hline
         \ce{Ba5Sr5(Co6O17)2} & (100) &  0.74 \\
         \ce{Ba5Sr5(CoO3)12} & (100) & 0.94 \\
         \ce{Ba3Sr3Co6O17} & (101) & 0.65 \\
         \ce{Ba5Ti12Sn5O34} & (100) & 0.45 \\
         \ce{Ba5Ti12Sn5O36} & (100) & 0.34 \\
         \ce{Ba3Ti6Sn3O17} & (101) & 2.20 \\
         \hline
    \end{tabular}
    \caption{De-prioritized surface energies in J/m$^{2}$.}
    \label{tab:discarted_surface_energies}
\end{table}

\section{Surface Pourbaix Diagrams}
\begin{figure}[H]
    \centering
    \includegraphics[width=0.9\textwidth]{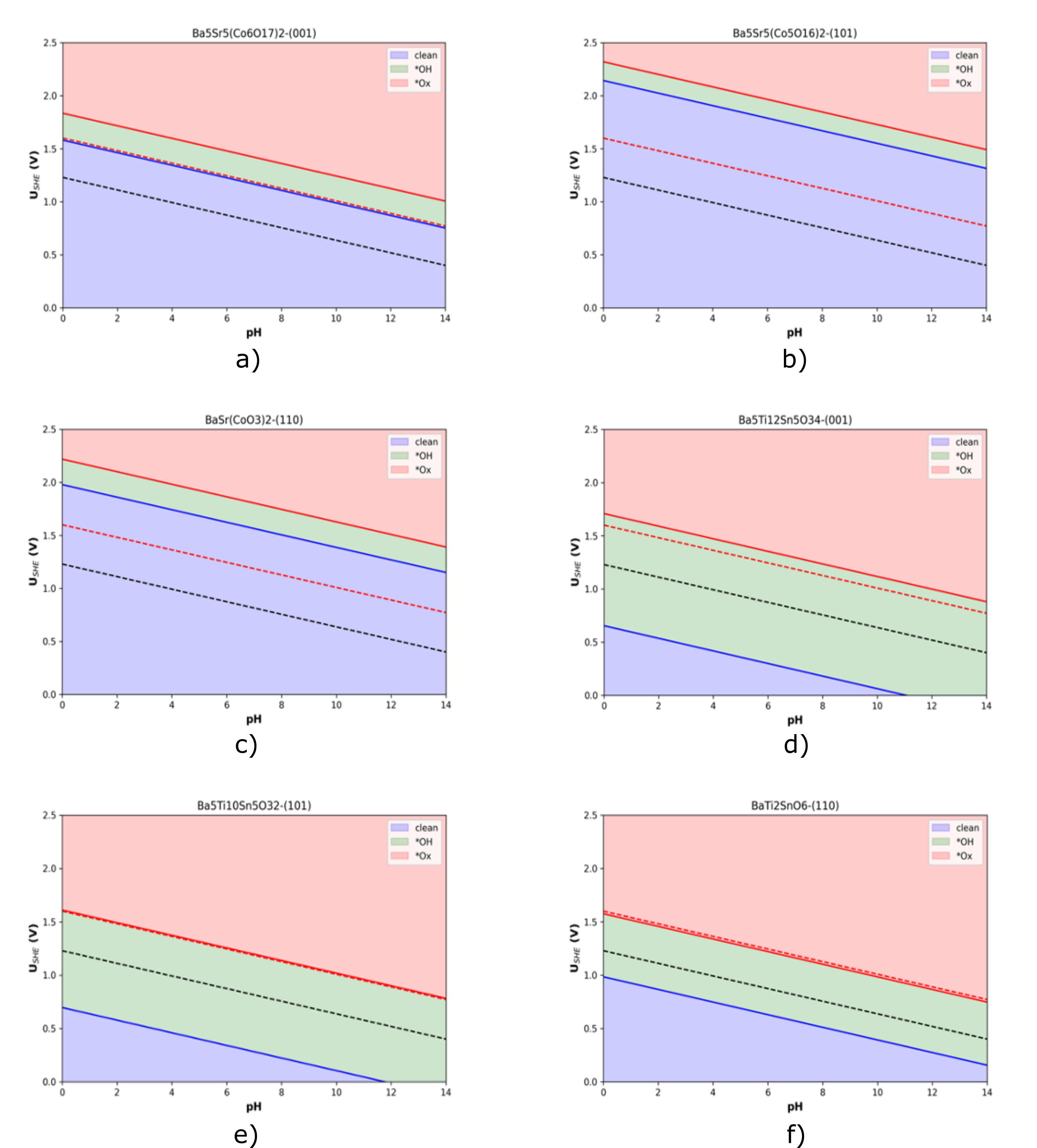}
    \caption{\small  Surface Pourbaix Diagrams across all materials and facets studied in this work.}
    \label{Surface_PBX_all_materials_facets}
\end{figure}
\section{DFT Figures \ce{BaSnTi_{2}O_{6}}}
\begin{figure}[H]
    \centering
    \includegraphics[width=0.8\textwidth]{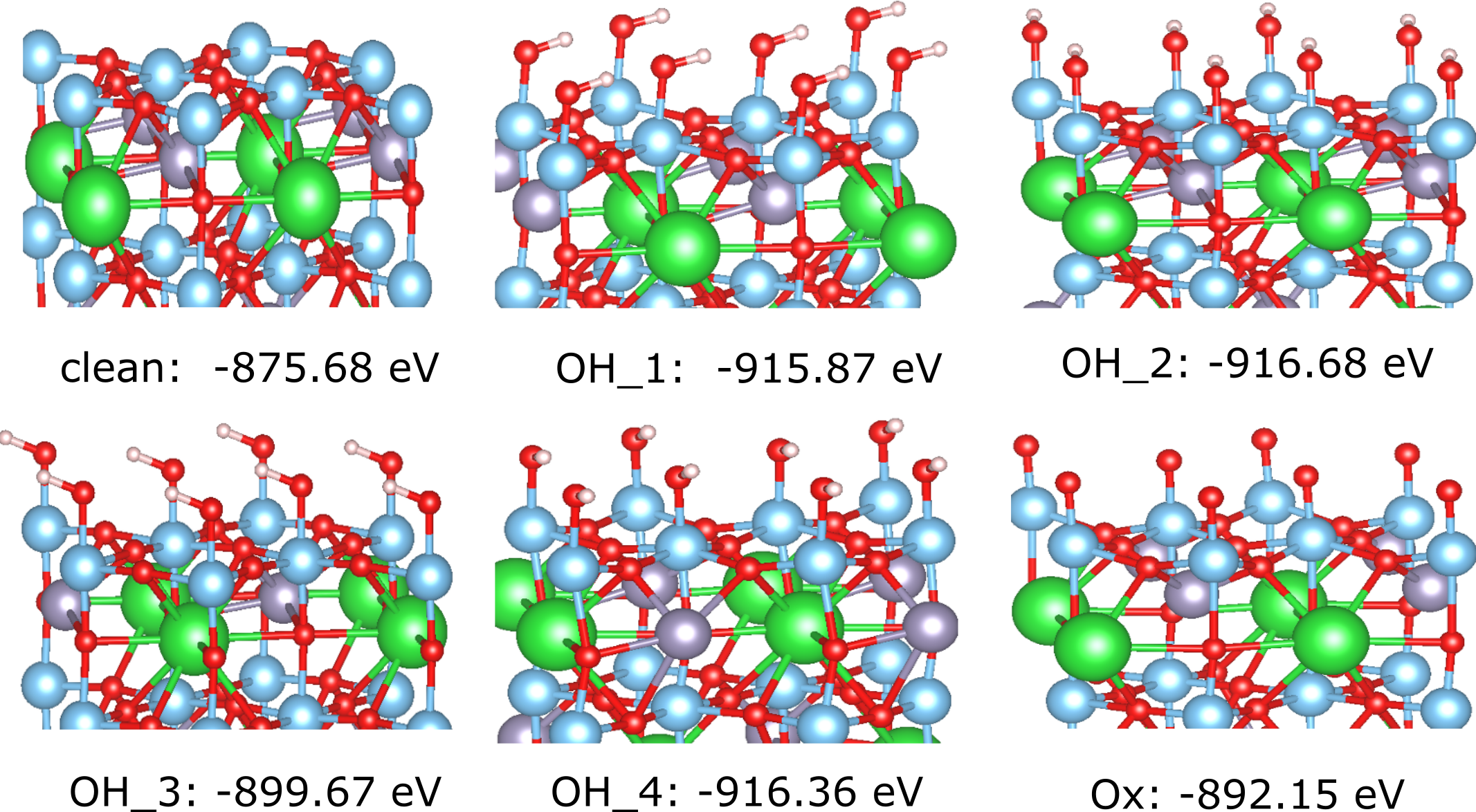}
    \caption{\small \ce{Ba5Ti12Sn5O34-(001)} surface Pourbaix Diagram Intermediates: From Clean (Top-Left), \ce{OH^{*}} rotation screening (1-4) and \ce{O^{*}}.}
    \label{BaSnTiSurfacePBX_001}
\end{figure}

\begin{figure}[H]
    \centering
    \includegraphics[width=0.8\textwidth]{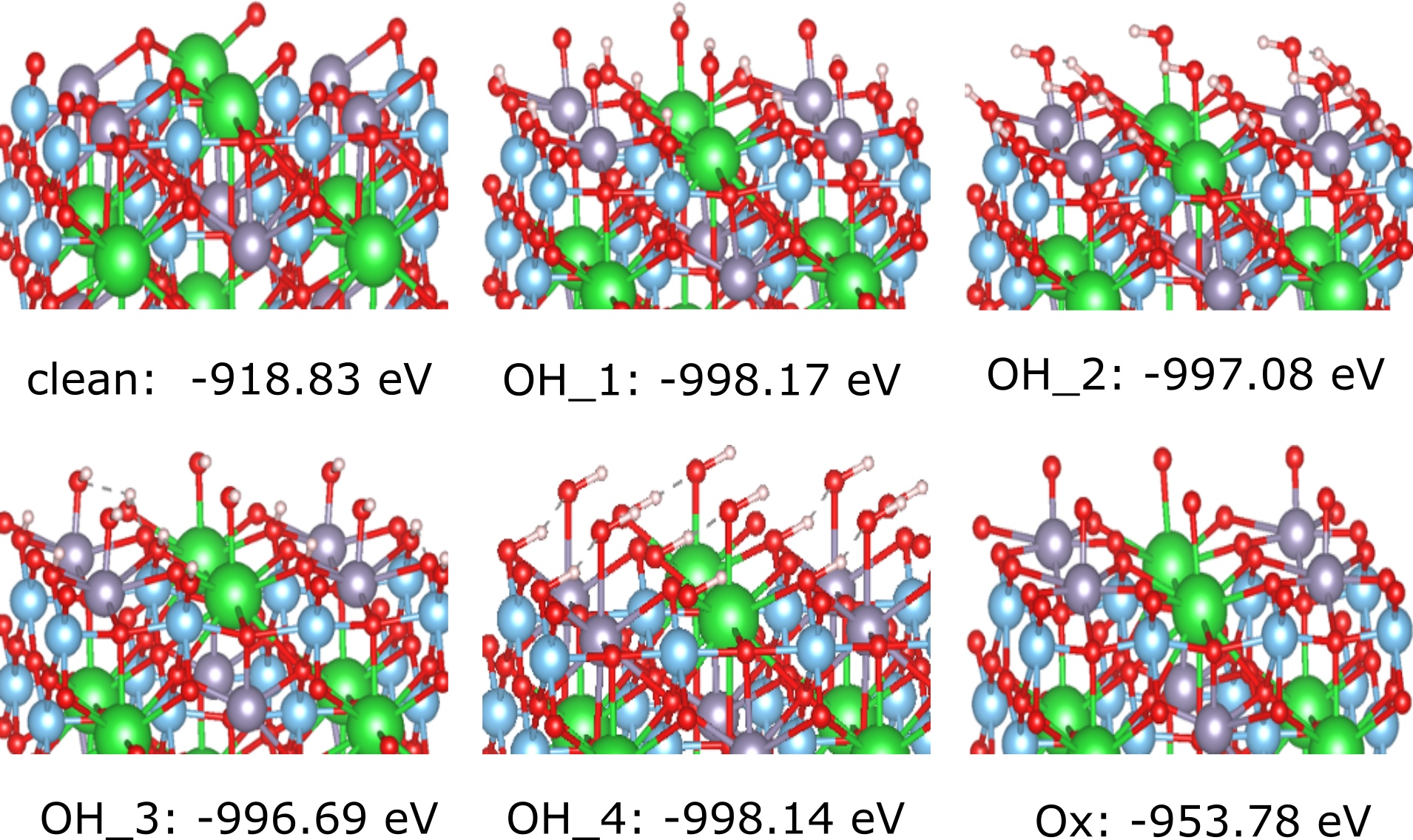}
    \caption{\small \ce{BaTi2SnO6-(110)} surface Pourbaix Diagram Intermediates: From Clean (Top-Left), \ce{OH^{*}} rotation screening (1-4) and \ce{O^{*}}.}
    \label{BaSnTiSurfacePBX_110}
\end{figure}

\begin{figure}[H]
    \centering
    \includegraphics[width=0.8\textwidth]{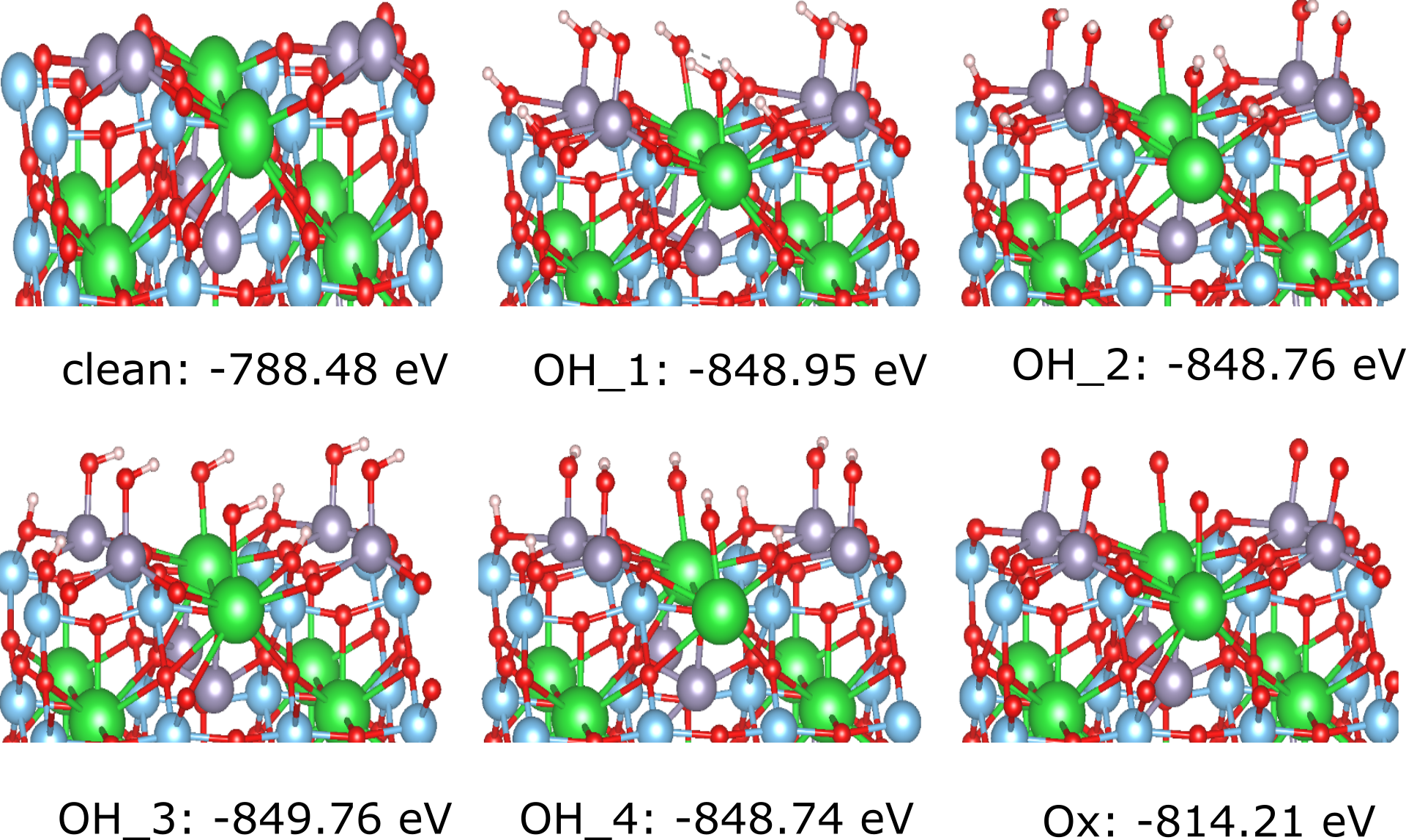}
    \caption{\small  \ce{Ba5Ti10Sn5O32-(101)} surface Pourbaix Diagram Intermediates: From Clean (Top-Left), \ce{OH^{*}} rotation screening (1-4) and \ce{O^{*}}.}
    \label{BaSnTiSurfacePBX_101}
\end{figure}

\begin{figure}[H]
    \centering
    \includegraphics[width=0.8\textwidth]{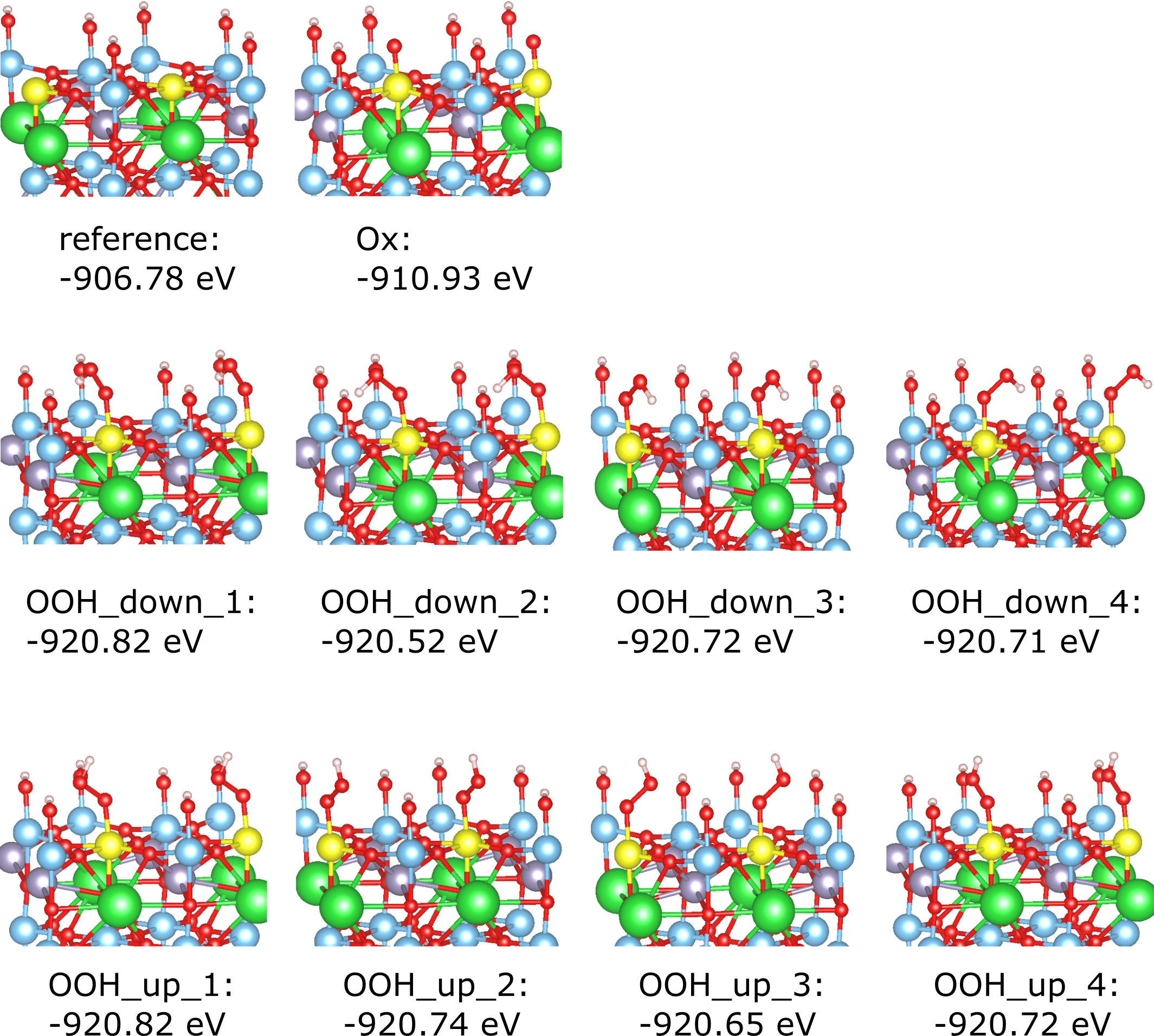}
    \caption{\small  \ce{BaSnTi_{2}O_{6}}  Reactivity for the *OH terminated (001).}
    \label{Reactivity_BaSnTiO_OH_001}
\end{figure}

\begin{figure}[H]
    \centering
    \includegraphics[width=0.95\textwidth]{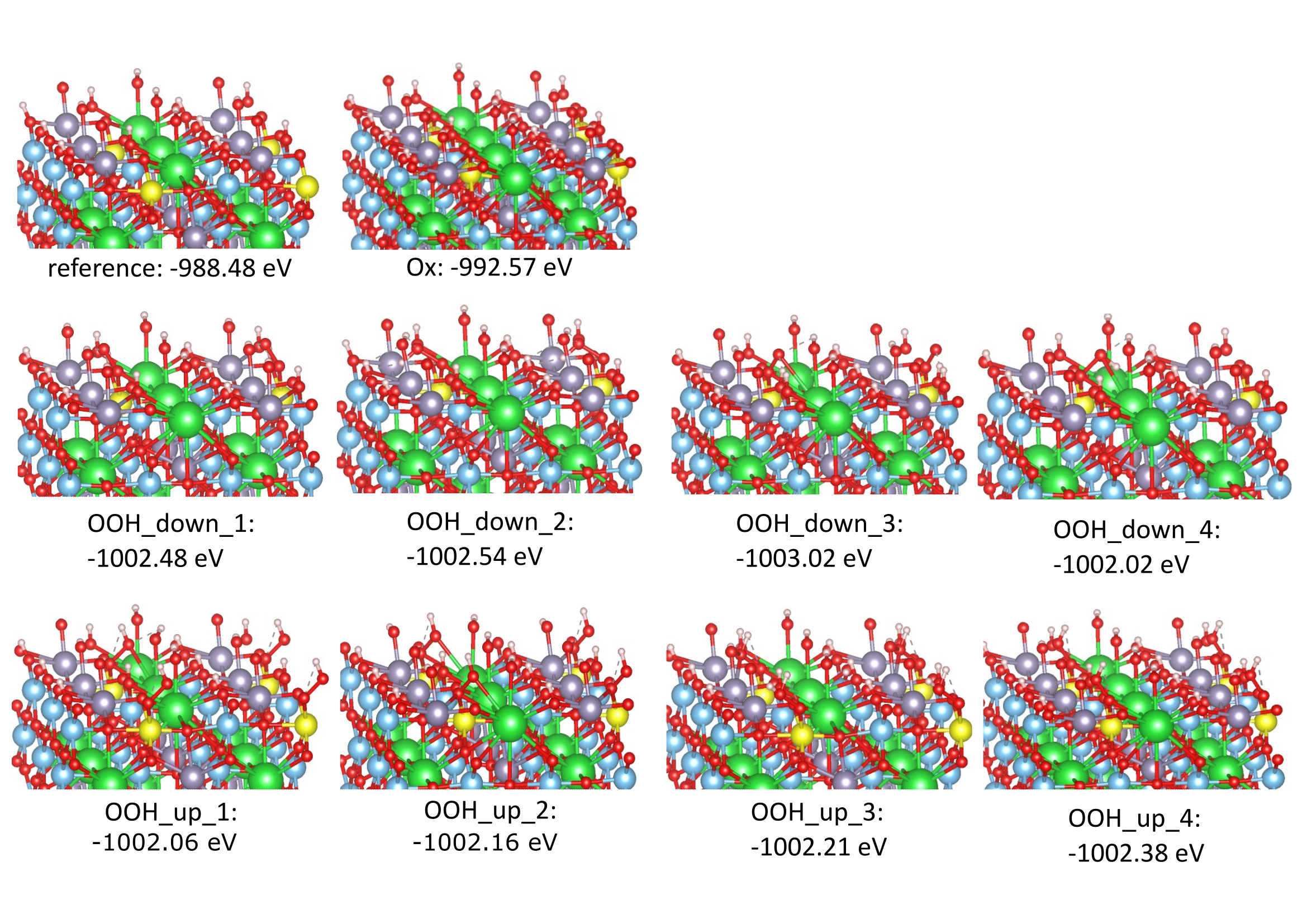}
    \caption{\small  \ce{BaSnTi_{2}O_{6}} Reactivity for the *OH terminated (110) surface.}
    \label{Reactivity_BaSnTiO_OH_110}
\end{figure}

\begin{figure}[H]
    \centering
    \includegraphics[width=0.8\textwidth]{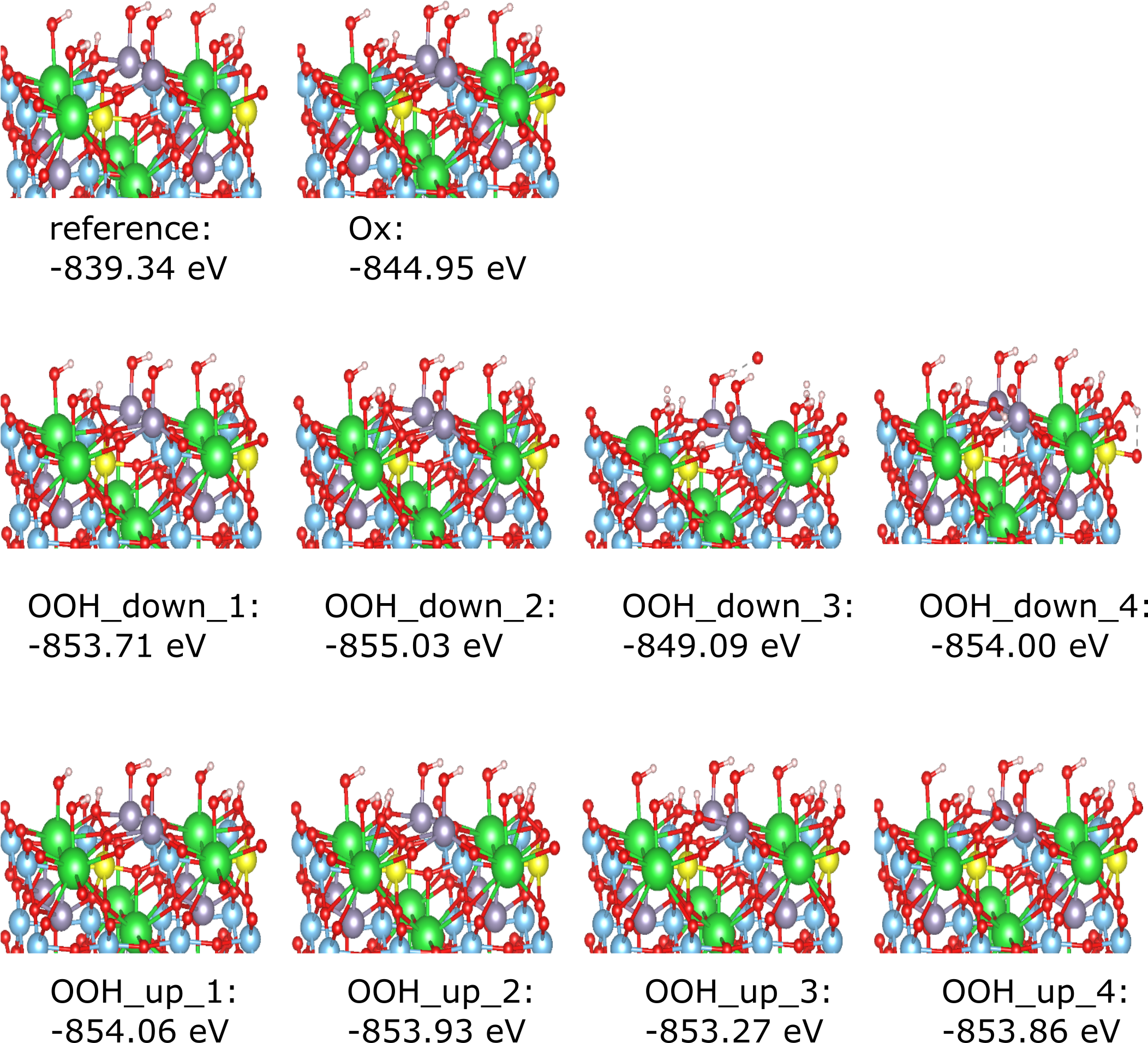}
    \caption{\small  \ce{BaSnTi_{2}O_{6}} Reactivity for the *OH terminated (101) surface.}
    \label{Reactivity_BaSnTiO_OH_101}
\end{figure}

\section{DFT Figures \ce{BaSrCo_{2}O_{6}}}

\begin{figure}[H]
    \centering
    \includegraphics[width=0.8\textwidth]{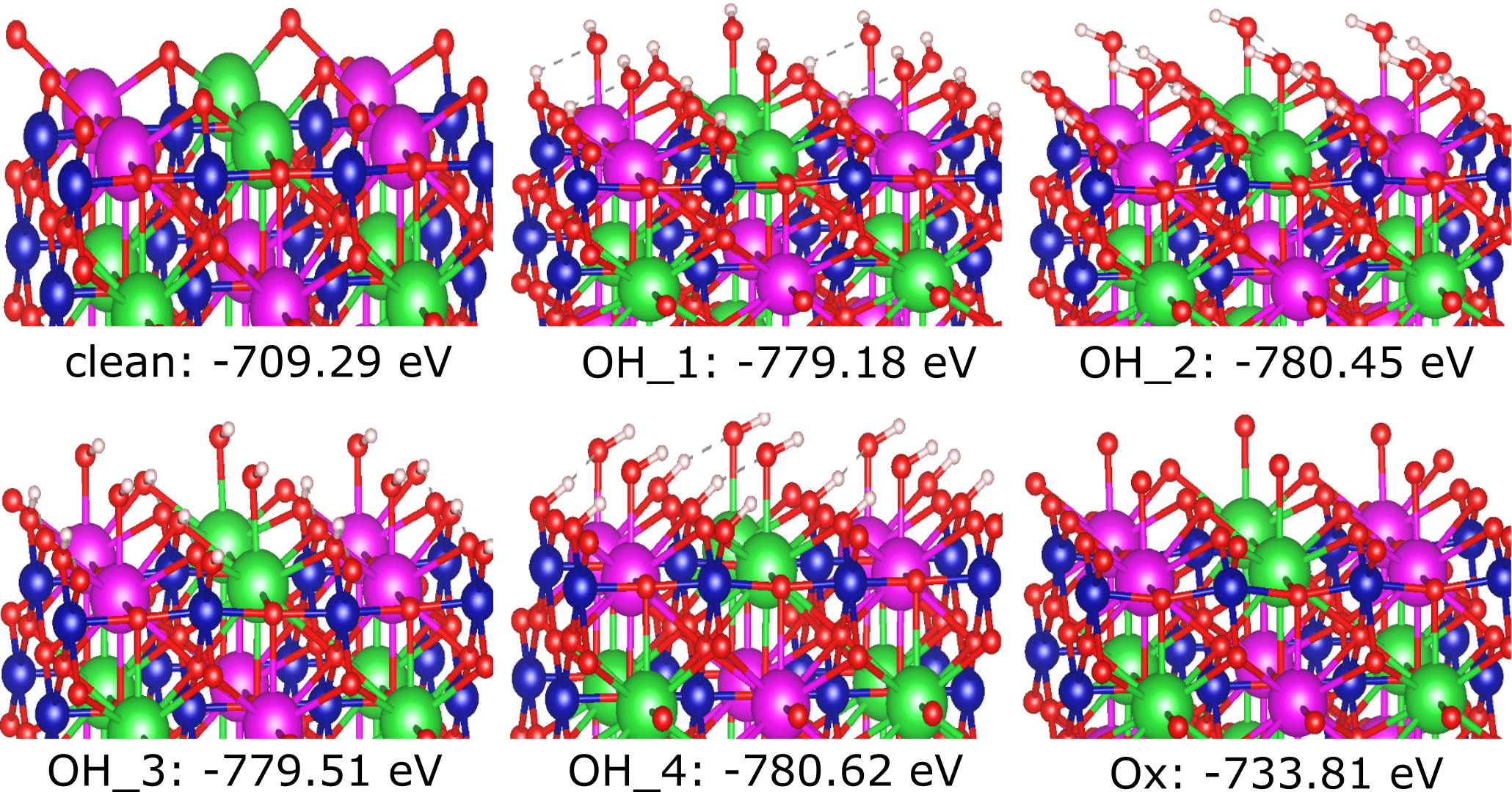}
    \caption{\small \ce{BaSr(CoO3)2-(110)} surface Pourbaix Diagram Intermediates: From Clean (Top-Left), \ce{OH^{*}} rotation screening (1-4) and \ce{O^{*}}.}
    \label{BaSrCoSurfacePBX_110}
\end{figure}

\begin{figure}[H]
    \centering
    \includegraphics[width=0.8\textwidth]{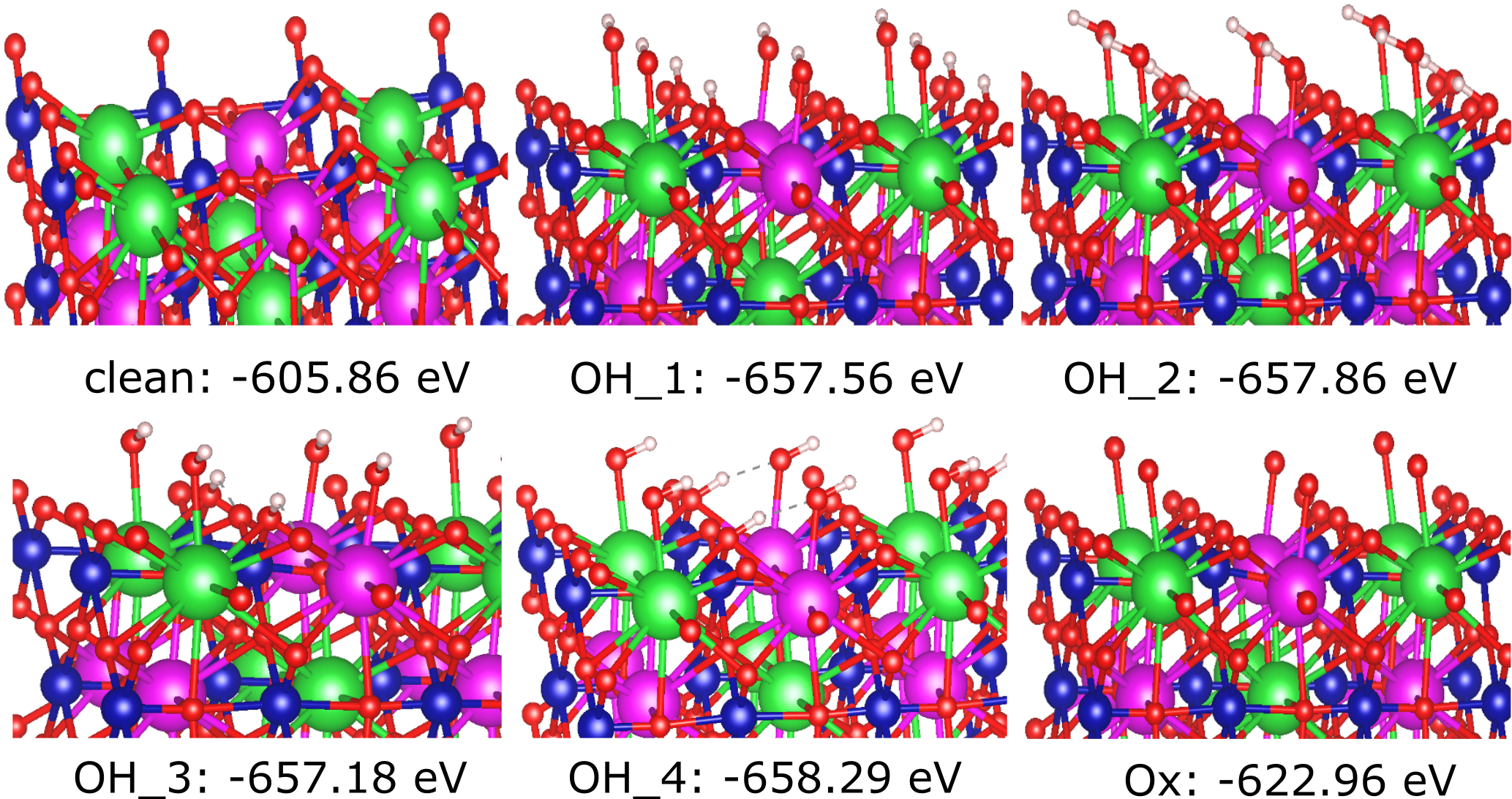}
    \caption{\small \ce{Ba5Sr5(Co5O16)2-(101)} surface Pourbaix Diagram Intermediates: From Clean (Top-Left), \ce{OH^{*}} rotation screening (1-4) and \ce{O^{*}}.}
    \label{BaSrCoSurfacePBX_101}
\end{figure}

\begin{figure}[H]
    \centering
    \includegraphics[width=0.8\textwidth]{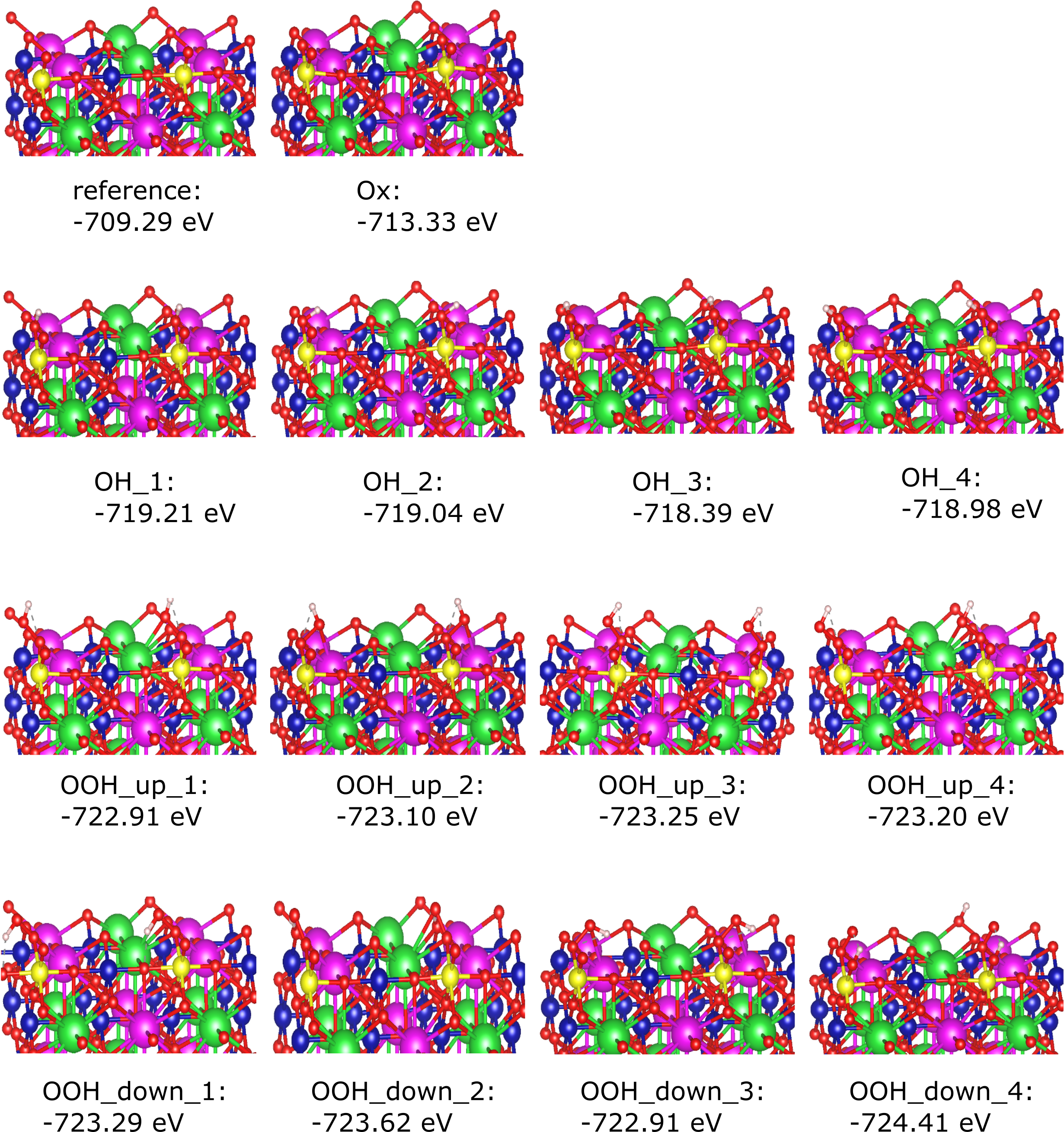}
    \caption{\small  \ce{BaSr(CoO3)2} Reactivity for the clean (110) termination.}
    \label{Reactivity_110}
\end{figure}

\begin{figure}[H]
    \centering
    \includegraphics[width=0.8\textwidth]{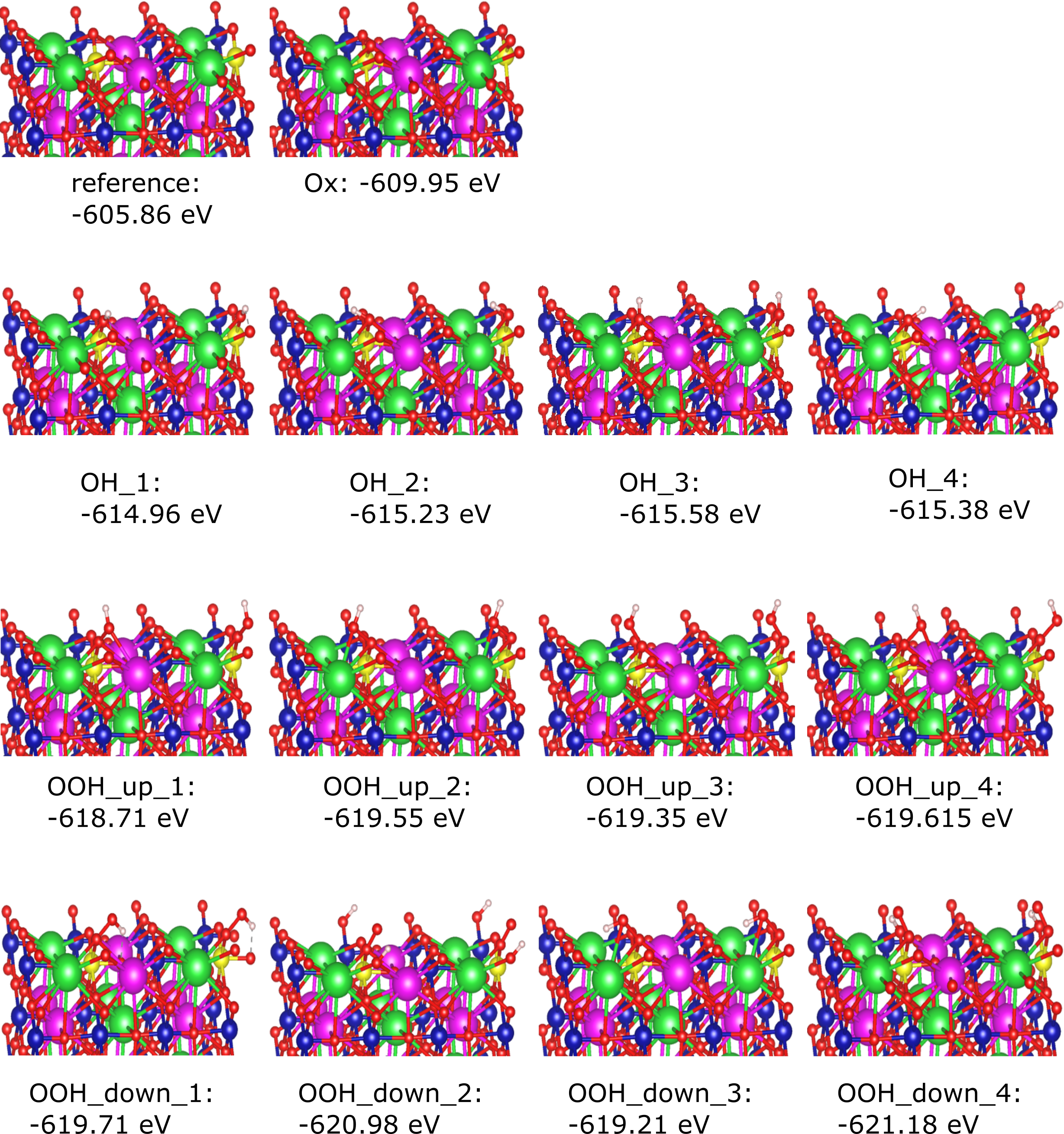}
    \caption{\small  \ce{BaSr(CoO3)2} Reactivity for the clean terminated (101).}
    \label{Reactivity_101}
\end{figure}
\clearpage
\twocolumn
\footnotesize
\bibliography{esi-reference.bib}

%% file: sections/abstract.tex
\begin{abstract}
    This paper introduces \textit{WhereWulff}, a semi-autonomous workflow for modeling the reactivity of catalyst surfaces. The workflow begins with a bulk optimization task that takes an initial bulk structure, and returns the optimized bulk geometry and magnetic state, including stability under reaction conditions. The stable bulk structure is the input to a surface chemistry task that enumerates surfaces up to a user-specified maximum Miller index, computes relaxed surface energies for those surfaces, and then prioritizes those for subsequent adsorption energy calculations based on their contribution to the Wulff construction shape. The workflow handles computational resource constraints such as limited wall-time as well as automated job submission and analysis. We illustrate the workflow for oxygen evolution (OER) intermediates on two double perovskites. \textit{WhereWulff} nearly halved the number of Density Functional Theory (DFT) calculations from $\sim$ 240 to $\sim$ 132 by prioritizing terminations, up to a maximum Miller index of 1, based on surface stability. Additionally, it automatically handled the 180 additional re-submission jobs required to successfully converge 120+ atoms systems under a 48-hour wall-time cluster constraint. There are four main use cases that we envision for \textit{WhereWulff}: \textbf{\em (1)} as a first-principles source of truth to validate and update a closed-loop self-sustaining materials discovery pipeline, \textbf{\em (2)} as a data generation tool, \textbf{\em (3)} as an educational tool, allowing users (e.g. experimentalists) unfamiliar with OER modeling to probe materials they might be interested in before doing further in-domain analyses, \textbf{\em (4)} and finally as a starting point for users to extend with reactions other than OER, as part of a collaborative software community.
\end{abstract}

%% file: sections/intro.tex
\section{Introduction}

One of the most challenging scientific problems of the 21$^{\text{st}}$ century is the development of sustainable technologies to produce, store and use clean energy\cite{seh2017combining, lewis2016research, montoya2017materials}. Renewable energy's intermittency (e.g. sunlight, wind or ties) requires efficient grid-scale storage to transfer power from times of excess generation to times of excess demand\cite{yang2011electrochemical}. To address this challenge, a number of promising storage techniques have been devised, one of which involves the storage of renewable energy into chemical bonds, e.g., water splitting to \ce{H2} or \ce{CO2} conversion to liquid fuels. With electrolyzer costs expected to fall by 60-80 \% in the next decade~\cite{ElectrolyzerCost}, hydrogen has evolved as an important medium for energy storage and  is projected to attract \$100 to \$150 billion investments by 2025~\cite{McKinsey}. However, the large-scale application of these technologies still relies on the availability of active, stable and cost-effective electrocatalysts for reactions like water splitting\cite{fabbri2014developments,reier2017electrocatalytic}, in which two water molecules evolve \ce{H2} and \ce{O2} gas. One major bottleneck of water splitting is the Oxygen Evolution Reaction (OER), hindering practical green hydrogen production due to slow kinetics, complicated bond rearrangements and the formation of an O-O bond. Although, state-of-the-art OER performances have been exhibited by ruthenium and iridium-based noble metal oxides\cite{mccrory2015benchmarking,yu2019recent}, the high cost and low abundance of these materials limits their practical application. Thus, it is paramount to design or search for alternative electrocatalysts with low-cost and earth abundant metals having catalytic performances comparable to the \ce{Ru}/\ce{Ir} benchmarks.

Transition metal oxides such as \ce{ABO3} perovskites have received significant attention for their environmental and energy-related applications\cite{labhasetwar2015perovskite,zheng2019pd,zheng2019cacoxzr1,hwang2017perovskites}, with OER activities rivaling that of \ce{IrO2} and \ce{RuO2}. The composition landscape of perovskites consists of a wide variety of elemental choices for the A and B sites, leading to a vast materials space. A class of perovskites for which this is apparent, is double perovskites (\ce{AA^{$'$}BB^{$'$}O_{6}}), where substitution of the B site with other transition metals has already been an effective approach to tune the d-band center to the Fermi level (E$_f$)\cite{suntivich2011perovskite}. Early efforts in materials design have hinged on chemical intuition and domain expertise, where strategies can often lead to incremental enhancement of existing material properties, rather than systematically searching the unexplored chemical space.

Recently, the catalysis community has witnessed the emergence of artificial intelligence in high-throughput materials design\cite{curtarolo2013high}. Machine learning algorithms offer the ability to establish the correlations between material structure and the properties of interest\cite{schmidt2019recent,schlexer2019machine}. Nowadays, machine learning has become one of the most attractive tools in materials research and specifically in catalysis. These advanced algorithms are applied to predict crystal properties such as formation energy, electronic properties, adsorption properties for surface chemistry and optical characteristics of metal oxides and organometallics\cite{xie2018crystal,chen2019graph,schutt2018schnet,klicpera2020fast,gasteiger2021gemnet}. Typically, materials data are available either in databases (e.g. Materials Project\cite{materials_project}, OQMD\cite{oqmd}) or can be generated through high-throughput computational approaches. For some classes of materials, acquiring a sufficient amount of unbiased materials data for model training is not always feasible, particularly for catalysis, in which the reactions occur at interfaces and defects. The main reason is that the description of the surface reactivity for those systems requires accurate quantum-chemical methods. Workflow engines are becoming crucial to address such challenges in computational materials design, providing fully automated computational task scheduling, high-scalability across distributed resources, data reusability, reliability and rapid prototyping.

Recent efforts embracing automated workflows have been undertaken for computational chemistry but also more specifically in the catalysis field. Notably, an automated adsorption energy workflow for semi-conductors~\cite{oxana}, where the authors provide a new and improved pipeline with minimal user supervision. Likewise, the automated bonding analysis with Crystal Orbital Hamilton Populations (COHP) workflow~\cite{george2022automated}, which enables high-throughput bonding analysis and facilitates the use of bonding information for machine learning studies, has been introduced. In this work, we couple deep expertise in quantum chemistry and catalysis with that in workflow engineering, echoing the thoughts of the recently published perspective~\cite{AEM}. We demonstrate both quantitatively and qualitatively, the benefits that arise from such synergy by conducting an automated and thorough analysis of two double perovskite materials: \ce{BaSrCo2O6} and \ce{BaSnTi2O6}, previously recommended by Zheng \textit{et al.} on the basis of their promising machine learning-predicted activity for OER. Our framework provides an autonomous bridge between the machine-learning assisted exploration of new chemical space and the update/enrichment of the training dataset with high quality DFT data~\cite{Hongliang}, usually referred to as ``closing the loop"~\cite{stein2019progress}. 

Our workflow, coined as \textit{WhereWulff}, addresses the following challenges in surface oxide computational chemistry modeling in catalysis: 1) the substantial compute time for conducting ab-initio calculations of multi-component surface oxides, exacerbated by the strong correlation effects and the large slab models required to enforce slab symmetry, 2) the extra degrees of freedom around magnetic moments on transition metals (TMs) in coordination environments, 3) the extra complexity around modeling the surface terminations at specific reaction conditions to ensure more representative adsorption energies of key intermediates.

In addition to tackling the aforementioned scientific challenges, the workflow streamlines and augments reactivity modeling with on-the-fly post-processing analyses such as surface energy calculations, Wulff shape and surface Pourbaix diagram constructions as well as reactivity pathway exploration. We also expect our open-sourced workflow to serve a didactic purpose, democratizing access to complex material science pipelines for experimentalists, who would like to corroborate or guide their endeavors but do not have the formal theoretical and computational training. Finally, in the same way that a community of scientists and software developers has helped deliver scientific software such as RosettaCommons~\cite{RosettaCommons}, we hope that by open-sourcing \textit{WhereWulff}, we can build a community to extend it with a spectrum of reactions. These extensions could be as simple as having to add new expressions for the theoretical overpotential, for instance, in the case of \ce{H2O2} production, where the intermediates of interest are also \ce{OH^{*}}, \ce{O^{*}} and \ce{OOH^{*}}.~\cite{h202}

%% file: sections/methods.tex
\section{Methods}

\subsection{Data and Software Availability}
 We leverage pre-existing open-source software packages, with the most noteworthy ones being Atomate~\cite{MATHEW2017140}, FireWorks~\cite{fireworks}, Pymatgen~\cite{custodian} and Custodian~\cite{materials_project,custodian}, in order to deliver our workflow, which is itself open-sourced at \repourl. The calculations that were carried out in this paper were all managed within the FireWorks framework, which advocates for Pilot Abstraction~\cite{pilot-abstraction}: the decoupling of job specification from resource allocation. A \textit{Firework} is an ensemble of tasks called \textit{Firetasks}, representing the smallest units of compute, customizable via Python. We conduct DFT calculations using the Vienna Ab initio Simulation Package (VASP)~\cite{kresse_furthmuller_1996,second_kresse}, the details of which are outlined in the following DFT Details section.

 \subsection{Overview}
\begin{figure}[H]
    \centering
    \includegraphics[width=0.90\textwidth]{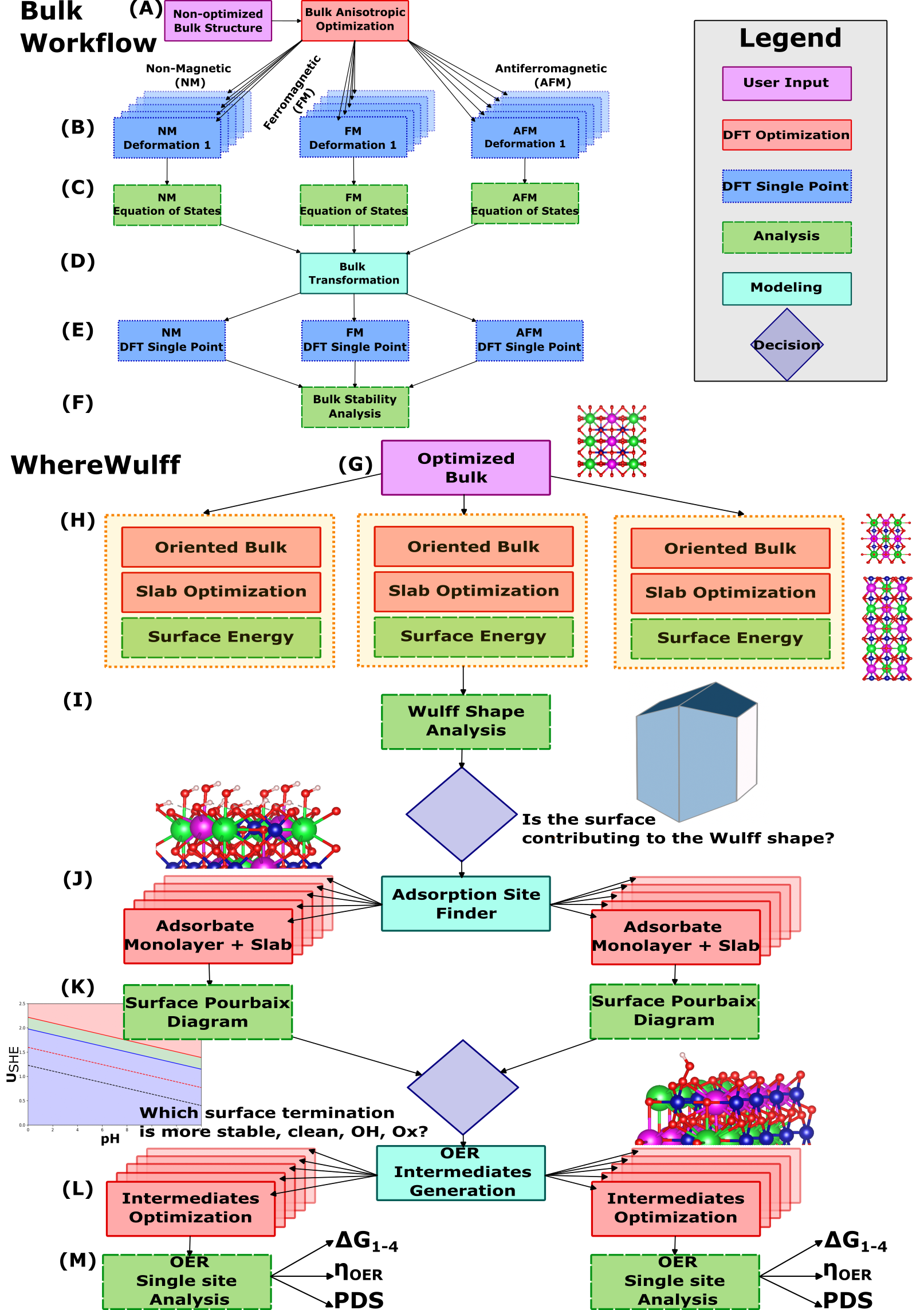}
    \caption{Schematic outlining the bulk optimization workflow, stages (A) through (F) and \textit{WhereWulff}, stages (G) through (M) and how they can be coupled via stage (G).}
    \label{Slab_workflow}
\end{figure}

 We provide two workflows (Figure~\ref{Slab_workflow}): an auxiliary bulk optimization workflow and our primary surface chemistry workflow, coined as \textit{WhereWulff}. While they can be used independently, the expectation is for a user to apply the bulk optimization workflow first to refine an unprobed material in terms of geometry, lattice parameters and magnetic ordering as well as to characterize that bulk material in terms of synthesizability and electrochemical stability prior to feeding it to \textit{WhereWulff}. \textit{WhereWulff} would then be used to initiate the surface cleavage, optimizations, surface energy ($\gamma_{hkl}$) characterizations as well as facet prioritization based on contribution to the Wulff construction shape. Subsequently, an end-to-end automated recipe from adsorbate placement, surface Pourbaix diagrams and reactivity modeling is implemented and applied per prioritized \onecolumn
\twocolumn \noindent facet. The very last step is the reactivity analysis for each facet, with the $\Delta G_{1-4}$, electrochemical theoretical overpotential ($\eta_{oer}$) and potential determining step (PDS) being stored as metadata in the database. We elaborate on both workflows in the following sections. In this particular work, we focus on OER but the workflow is modular and flexible enough to include other types of reaction mechanisms. 
 
 \subsection{Bulk Optimization Workflow}
 The first \textit{Firework} in our bulk optimization workflow conducts an anisotropic optimization of the bulk crystal structure (Figure~\ref{Slab_workflow}-A), the results of which are stored in the database as a crystal template for the following steps. Magnetic properties are key to understanding and properly representing the electronic structure of any material, especially when it comes to multi-metallic oxides. As a result, the subsequent crucial step in the bulk workflow involves a magnetic configuration search across non-magnetic (NM), ferromagnetic (FM) and antiferromagnetic (AFM) orderings, to determine the most stable state in terms of lattice parameters, atomic degrees of freedom and magnetic configuration of the given bulk structure. This search is carried out through parallel isotropic transformations and single point calculations in order to build the equation of states (EOS)\cite{latimer2018evaluation, vinet1987compressibility} fit for the three magnetic configurations (Figure~\ref{Slab_workflow}-B). The magnitude of the atomic magnetic moments are derived by decorating the bulk structure with the corresponding oxidation states and following the Crystal Field Theory (CFT)\cite{mabbs2008magnetism}, which describes the splitting of degenerate orbitals in cations surrounded by anion charges. After the EOS stage (Figure~\ref{Slab_workflow}-C) is completed, the equilibrium lattice parameters are extracted, for each ordering, and the template bulk structure transformed (Figure~\ref{Slab_workflow}-D) with a view to computing the DFT energy for each of the three configurations (Figure~\ref{Slab_workflow}-E). The most stable bulk structure is then chosen based on the DFT energy and a post-processing analysis \textit{Firework} that automatically builds the phase diagram and bulk Pourbaix diagram (Figure~\ref{Slab_workflow}-F). Those automatically extract the energy above the hull (E$_{hull}$) and the delta Gibbs of decomposition ($\Delta G_{pbx}$) at a given pH and voltage, characterizing the bulk material's synthesizability and electrochemical stability respectively. All this metadata can be easily queried as part of a hosted database and ultimately fed into \textit{WhereWulff} for probing the bulk material's surface properties.
\begin{figure}[H] 
    \centering
    \includegraphics[width=0.45\textwidth]{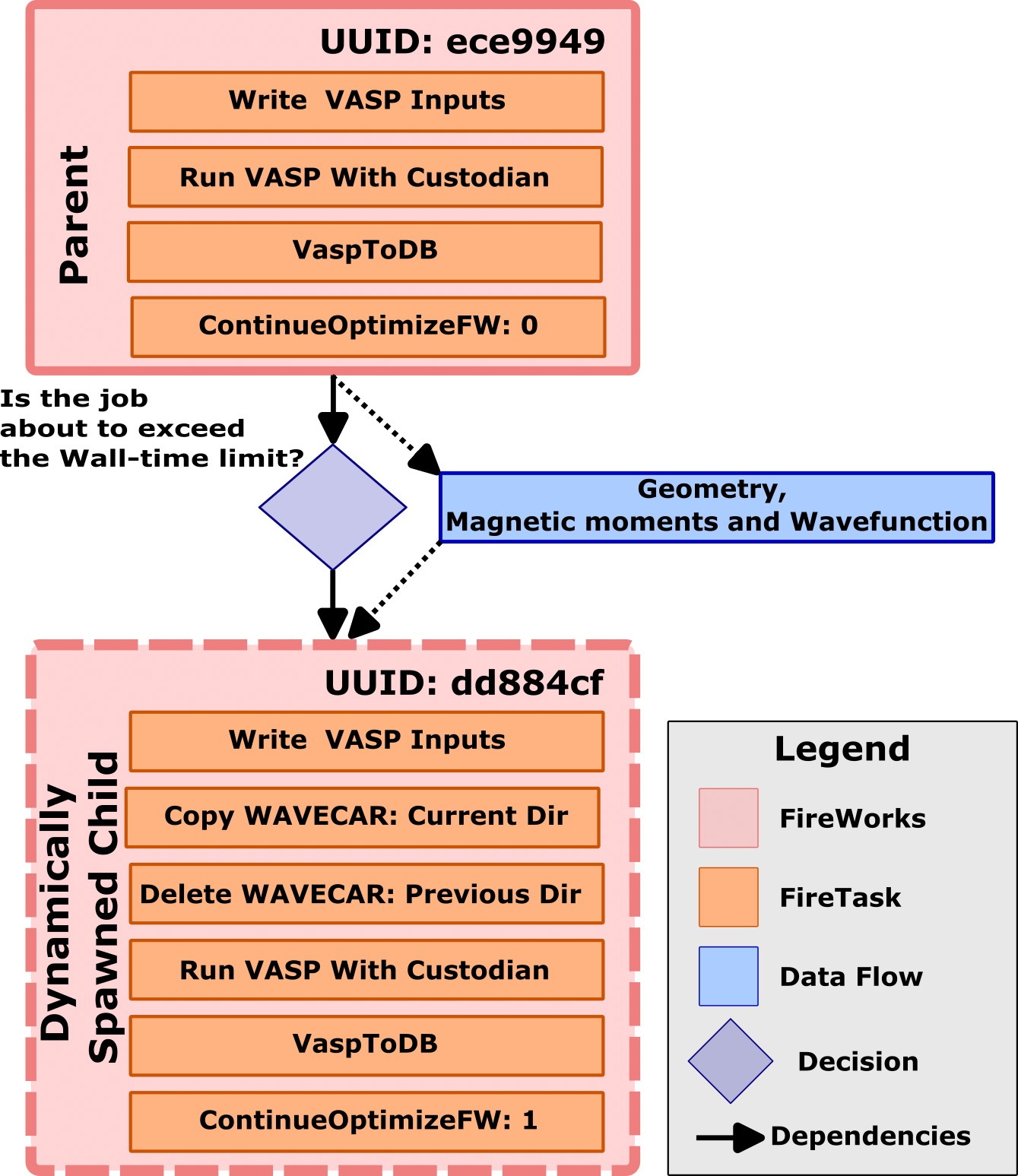}
    \caption{Schematic illustrating the \textbf{\textit{ContinueOptimizeFW}} protocol, through which long-running optimization jobs are continued in a way that simultaneously unlocks scale, provenance and efficiency. The scale comes from a Custodian~\cite{custodian} daemon, which autonomously triggers a clean termination of the VASP processes based on a specified wall-time and the running average time taken for one iteration of VASP. The provenance comes from a dynamic mutation of the workflow graph through Firework's~\cite{fireworks} \textit{detours} functionality. Finally, the efficiency is attributed to \textit{WhereWulff}'s robust message passing scheme. The latter uses UUIDs to safely mutate and retrieve atomistic metadata (geometry, magnetic moments and wavefunction) from a hosted MongoDB as well as to relay that metadata between the parent and child \textit{Fireworks}, allowing the child to start from a state that most closely matches the last state in the cleanly terminated parent Firework.}
    \label{continue_opt}
\end{figure}
\subsection{WhereWulff}
The only manual user intervention with \textit{WhereWulff} (Figure \ref{Slab_workflow}-G) is the provision of an equilibrium bulk structure for a given material, with the workflow handling the rest of the downstream probing, filtering and analysis toward the identification of stable and active facets for OER. From then on, \textit{WhereWulff} takes over, enumerating and cleaving a set of slab models for each symmetrically distinct Miller index depending on the maximum number of Miller indices specified by the user, enforcing the creation of symmetric slab models and as a result non-dipolar slabs. \textit{WhereWulff} manages each of these slab models as \textit{Fireworks} made up of three \textit{Firetasks}, each of which is tagged with a universally unique identifier (UUID), allowing one to probe the same material multiple times without database conflicts.  As soon as the two parent \textit{Firetask} optimizations are completed, the surface energy \textit{Firetask}, complying with dependency constraints, calculates the surface energy by using the chemical potential of the excess or lack of a given atomic species in the slab formula, following the work of Reuter \textit{et al.}~\cite{non-stoich} (see section S1). A custom \textbf{\textit{ContinueOptimizeFW}} (Figure~\ref{continue_opt}) \textit{Firetask} triggers \textit{detours}, which is a modification to an original workflow graph that inserts additional \textit{Fireworks} or \textit{Workflows}, to dynamically handle wall-time constraints. Additionally, this same \textit{Firetask}, upon confirming the successful convergence of an optimization, relays the relevant UUIDs to the downstream \textit{Fireworks} as part of its message passing scheme. Post slab model optimizations and surface energy calculations (Figure \ref{Slab_workflow}-H), \textit{WhereWulff} collects these UUIDs, which it uses to retrieve the correct surface energies and accompanying metadata from the database. With the correct information, \textit{WhereWulff} then converges to the next post-processing task, Wulff construction analysis\cite{wulff1901frage}, shown in Figure \ref{Slab_workflow}-I. \textit{WhereWulff} uses the Wulff shape construction as a way to prioritize the surfaces that contribute significantly to the nanoparticle shape, in terms of space group symmetry of the bulk and the surface energies, providing insight into which facets are most likely to be observed experimentally.

At this stage (Figure~\ref{Slab_workflow}-I), \textit{WhereWulff} exhibits another case of decision-making ability: it prioritizes the slab models that show the highest contribution to the nanoparticle shape. For each of the prioritized slab models, \textit{WhereWulff} leverages the adsorption site finder~\cite{OC22} (Figure~\ref{Slab_workflow}-J) to identify potential adsorption sites on the clean surface. This is accomplished by exploiting the Wyckoff and equivalent positions in the bulk and overlaying them onto the surface to locate the exposed TMs at the interface based on a decrease in the coordination number. \textit{WhereWulff} then places a set of user-defined and decorated adsorbates, which in this case consists of \ce{OH^{*}} and \ce{O^{*}}, to create extreme monolayers. It is important to mention that while this algorithm is applied to a pristine surface, the adsorption sites are then automatically adjusted to account for the structural relaxation of that surface. The optimized surface acts as a good initial geometrical guess for the subsequent termination relaxations. For polyatomic adsorbates like \ce{OH^{*}}, \textit{WhereWulff} also performs a configuration search, performing 360$^{\circ}$ rotations, about the \textit{z-axis}, in increments of 90$^{\circ}$, to account for the potential stabilization effects of hydrogen bonds between the adsorbate and its local surface environment. After orchestrating the set of adslab monolayer optimizations, \textit{WhereWulff}'s surface Pourbaix analysis \textit{Firetask} collects, by means of relayed UUIDs, the corresponding adslab energies, which it uses to construct a surface Pourbaix plot with three boundaries: clean, \ce{OH^*} and \ce{O^{*}} (Figure~\ref{Slab_workflow}-K). The task then applies the user-defined reaction condition to automatically come up with the most stable of the three surface terminations (clean, \ce{OH^*} and \ce{O^{*}}), which is relayed to the downstream reactivity tasks.

At this point, \textit{WhereWulff} has all the information to perform its reactivity analysis on the most stable surface termination (Figure \ref{Slab_workflow}-L). It randomly selects an active site based on a user-defined metal type to perform the catalytic process. If the surface coverage is either \ce{OH^{*}} or \ce{O^{*}}, \textit{WhereWulff} will skip and retrieve those results automatically, focusing on the OER intermediates that are yet to be calculated. This is achieved by placing the \ce{OH^{*}}/\ce{O^{*}} and the \ce{OOH^{*}}. The \ce{OOH^{*}} is specific to the Water Nucleophilic Attack (WNA) mechanism, and also posseses rotational degrees of freedom. These are factored in by rotating the adsorbate in 90$^{\circ}$ increments along the \textit{z-axis} and also by including two separate configurations, which differ in  whether the hydrogen atom is pointing towards the surface or toward the interlayer space. This allows us to automatically search for the lowest energy conformation, which is usually the one with the strongest hydrogen bond between the \ce{OOH^{*}} adsorbate and the nearest oxygen atom. Finally, all OER intermediates for all the slab models contributing to the nanoparticle shape decorated with the most stable termination are analyzed by the last post-processing analysis (Figure \ref{Slab_workflow}-M), which derives the $\Delta G_{1-4}$, the potential determining step (PDS) and the theoretical overpotential ($\eta_{oer}$) for each termination (see section S2) and stores this information in a database that can be queried.
\subsection{Deployment and regression tests}
In the vein of software best practices, we have set up an automated regression test to make sure that any new features do not break existing functionality. Using atomate's~\cite{MATHEW2017140} \textbf{\textit{RunVaspFake}} \textit{Firetask}, we are able to simulate the execution of VASP in a feasible time-frame for testing. We have set up the test, which is an end-to-end execution of the workflow on \ce{IrO2} via \textit{GitHub Actions}, such that it gets triggered every time someone wants to merge a new feature into the \textit{main} branch. Any time a deviation from the expected outputs from running the workflow on \ce{IrO2} is observed, for instance, the surface energy outputs differ from what was previously asserted, a failure notification is triggered. 

It is also important to mention that while the original purpose of \textbf{\textit{RunVaspFake}} was to allow our regression tests to run in a reasonable time, it has also allowed us to submit what we coin as \textit{hybrid} workflows: a mixture of workflow nodes that simulate VASP and others that run the actual binary, thereby allowing us to re-start a workflow at a certain point in the graph, without incurring the significant previous compute and with the ability to change the ensuing nodes in the workflow with a new feature or a bug fix.

\subsection{DFT Details}
All DFT periodic boundary calculations were performed within the spin-polarized formalism as implemented in the Vienna Ab-initio Simulation Package (VASP-6.2.1), and using the GGA PBE functional\cite{kresse_furthmuller_1996,second_kresse}. Ionic cores were described with the projector augmented wave (PAW) pseudopotentials\cite{blochl1994projector,kresse1999ultrasoft} and the valence electrons were represented through a plane-wave basis set with a kinetic energy cut-off of 500 eV. A (50/a, 50/b, 50/c) and (30/a, 30/b, 1) $\Gamma$-centered k-point mesh\cite{monkhorst1976special} was employed to describe the first Brillouin zone for bulks and surfaces, respectively. These settings were found to provide a good compromise between accuracy and computational time as illustrated in the convergence tests in section S3 of the SI. The energy convergence criterion was fixed to 10$^{-4}$ eV for the electronic structure, while the Hellman-Feynman forces criterion for geometry relaxation was set to 0.05 eV $\angstrom^{-1}$. These specifications were implemented as part of a custom class inheriting from the \textit{MVLSlabSet} input set class of \textit{Pymatgen}\cite{custodian}. As a result, we are able to standardize, version control and memorialize the VASP parameters across our various clusters and jobs. The rotationally invariant implementation of the Hubbard-U model by Dudarev~\cite{dudarev} was employed and applied to the 3d electrons of Co atoms to account for strong electron correlation effects. The slab model is made up of a 2D surface and a corresponding oriented bulk, since this has been shown to most efficiently converge the surface energy calculations~\cite{sun_ceder_2013}. As part of non-stoichiometric surface energy calculations, a metal species was consistently chosen as the reference species against which to compute free energy excess, per Eq. (\ref{bulk_energy_formula_unit}),
\begin{equation}
    \gamma = \frac{\bigl[G_{\text{slab}} - N_{\text{r}}g_{\text{bulk}} - \sum_{i}\bigl(N_i - x_{i} N_{\text{{r}}}\bigr) \mu_{i}\bigr]}{2A}
    \label{bulk_energy_formula_unit}
\end{equation}
 where $G_{\text{slab}}$ is the free energy of the slab, $g_{\text{bulk}}$ the bulk energy of the oriented unit cell and $A$ the cross-sectional area of a symmmetric slab. The last sum represents the free energy excess, with $x_{i}$ being the number of atoms per bulk formula and $N_\text{r}$ the reference specie that is picked. As a corollary, stoichiometric slabs are always independent of chemical potentials. Additionally, we neglect the ZPE and entropy corrections for $G_{\text{slab}}$ and $g_{\text{bulk}}$, which allows us to use the DFT energies, $E^{DFT}_{\text{slab}}$ and $E^{DFT}_{\text{bulk}}$ respectively.

\subsection{Other Parameters}
While \textit{WhereWulff} was built with minimal user intervention in mind, its interface is flexible enough 
for experienced users to change some of the default behaviour, as depicted in purple in Figure \ref{Slab_workflow}-A and G.
When it comes to the bulk optimization workflow, the user is required to provide the bulk structure of interest as well as the number of isotropic \textit{deformations}, to be applied toward the construction of the equation of states (EOS), the VASP \textit{magmom buffer} and whether or not to convert to the conventional standard bulk structure. Similarly, \textit{WhereWulff} can also be customized by the maximum number for (hkl) \textit{Miller indices}, whether or not to \textit{symmetrize} the slab model, perform \textit{slab repetition} at the cross-section, 
 and include \textit{selective dynamics}. The user can also provide the adsorbates list, the applied potential and pH to apply to the surface Pourbaix diagram as well as control the metal site on which to perform the reactivity analysis. The default values have been chosen based on benchmarking and convergence tests (section S3). 

%% file: sections/results.tex
\section{Case Study: Double Perovskites}
\begin{figure*}[ht]
    \centering
    \includegraphics[width=0.7\textwidth]{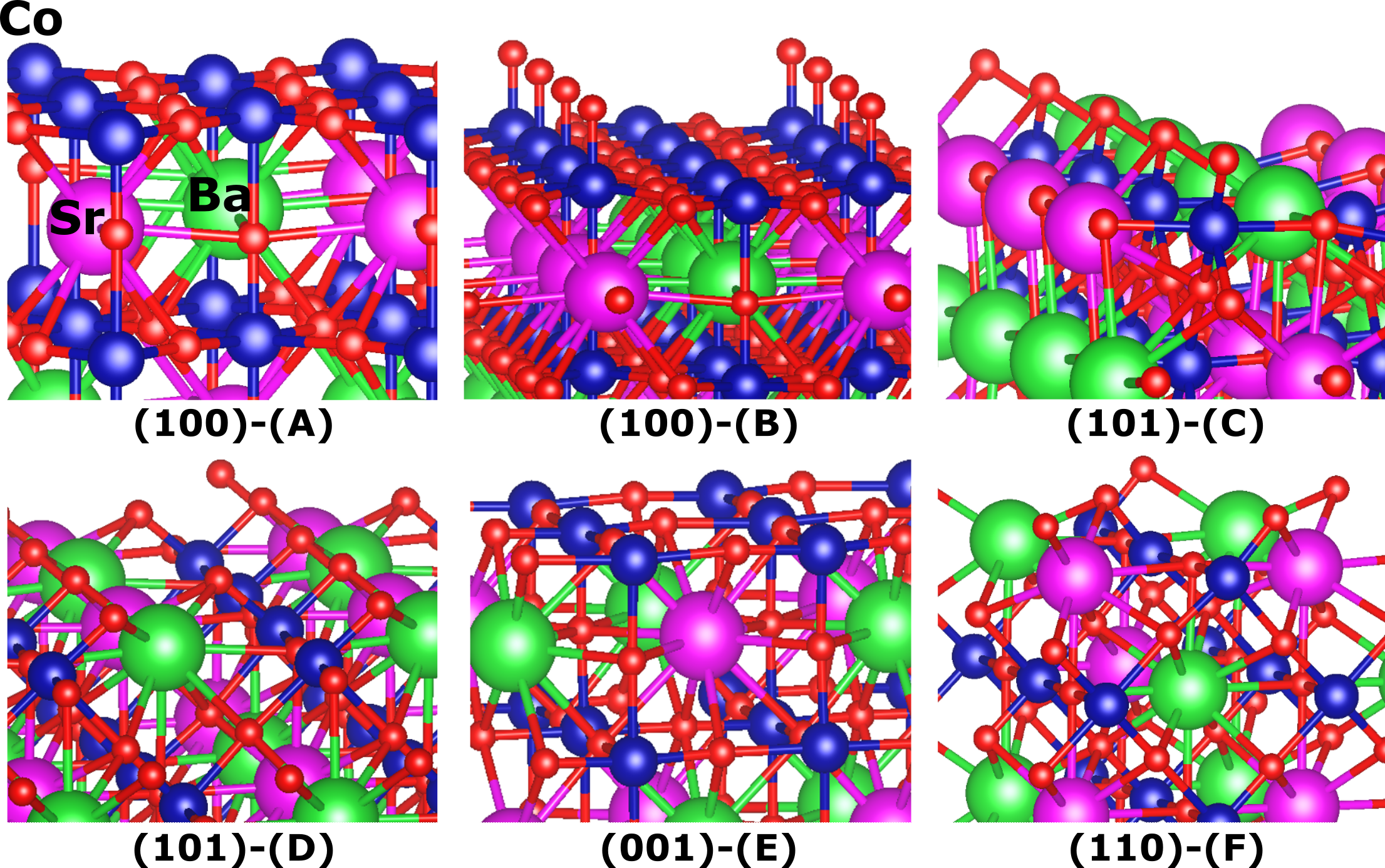}
    \caption{The six relaxed low Miller-index terminations probed by \textit{WhereWulff} for the two perovskites that were studied. \ce{Co} is represented in blue, \ce{Ba} is represented in green, \ce{Sr} is represented in purple and \ce{O} is depicted in red.}
    \label{clean_term_BaSrCo}
\end{figure*}

To showcase the workflow's value proposition as a potential first-principles source of truth in the context of a self-contained cycle with high-throughout machine learning-assisted exploration, we performed a thorough analysis of two of the previously unprobed double perovskites that were suggested by Zheng \textit{et al.}'s~\cite{Hongliang} machine learning pipeline. As shown in Figure~\ref{clean_term_BaSrCo}, \textit{WhereWulff} probed six different terminations across four symmetrically distinct low Miller indices and two materials. \ce{Ba5Sr5(Co6O17)2-(100)} facet exhibits exposed \ce{Co_{5c}} cations, with all of them undercoordinated, with the adsorption sites lying along the (hkl) direction (Figure \ref{clean_term_BaSrCo}-A), \ce{Ba5Sr5(CoO3)12-(100)} facet shows a mixture of fully-coordinated \ce{Co_{6c}} cations and exposed \ce{Co_{5c}} cations, with adsorption sites lying along the surface normal (Figure \ref{clean_term_BaSrCo}-B). \ce{Ba3Sr3Co6O17-(101)} facet shows undercoordinated \ce{Co_{5c}} cations (Figure \ref{clean_term_BaSrCo}-C).
Unlike for the (001) and (100) orientations, these sites are shared between the alkali metals and the transition metal. \ce{Ba5Sr5(Co5O16)2-(101)} facet exhibits a mixture of fully-coordinated \ce{Co_{6c}} cations and exposed \ce{Co_{5c}} cations, with adsorption sites shared between the transition metal and the alkali metals (Figure \ref{clean_term_BaSrCo}-D). \ce{Ba5Sr5(Co6O17)2-(001)} termination shows fully undercoordinated surface \ce{Co_{5c}} cations (Figure \ref{clean_term_BaSrCo}-E). \ce{BaSr(CoO3)2-(110)} is the only stoichiometric termination, where all the surface \ce{Co_{5c}} cations are undercoordinated, with the adsorption sites lying at an angle to the surface normal (Figure \ref{clean_term_BaSrCo}-F). As part of its autonomous probing, \textit{WhereWulff} identified 3 terminations (110), (101) and (001) for both \ce{BaSrCo2O6} and \ce{BaSnTi2O6}, on the basis of its Wulff construction. The other six terminations were de-prioritized based on a threshold contribution to the Wulff nanoparticle shape of at least 10 \%. This filter, shown in Figure~\ref{Slab_workflow}-I, saved us from having to perform at least 108 DFT calculations for surfaces that are not likely to contribute to the nanoparticle shape experimentally and explains the origins of the workflow's name: \textit{WhereWulff} guides the downstream activity analyses based on stability metadata it generates upstream. Details on the compositions for the de-prioritized surfaces and their surface energies are provided in the supporting information (section S4). Surface Pourbaix diagrams for the prioritized terminations were generated and are shown in section S5. Based on alkaline reaction conditions, the most stable coverage (clean, OH*, O*) was picked and the single-site OER reactivity downstream tasks were spawned. 

\begin{figure}[H]
    \centering
    \includegraphics[width=0.8\textwidth]{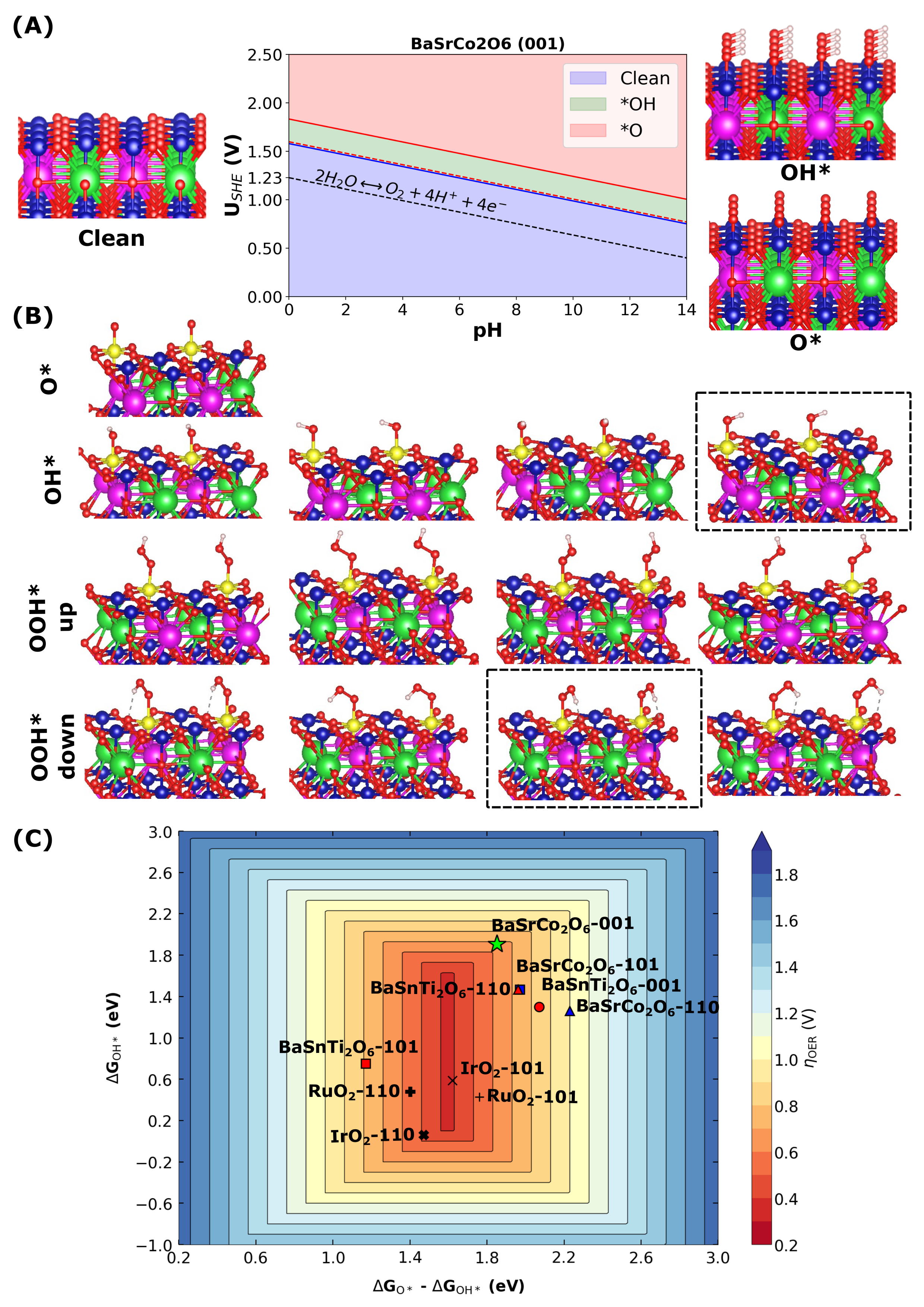}
    \caption{\small Shows the benchmark, \ce{BaSrCo2O6-(001)} (green star), whose OER activity we compared with results from Zheng \textit{et al.}~\cite{Hongliang}. \textbf{(A)} Result from the surface Pourbaix analysis part of \textit{WhereWulff} in Figure~\ref{Slab_workflow}-E, which identifies the most stable coverage under alkaline conditions, defined as the surface coverage that is the most expressed in the applied potential range of 1.23 $V_{RHE}$ to 1.6 $V_{RHE}$. \textbf{(B)} Results from the reactivity part of \textit{WhereWulff} in Figure~\ref{Slab_workflow}-F. The yellow colored atom is a Co active site. The black dashed boxes highlight the most stable configuration. \textbf{(C)} Two-dimensional OER activity plot of theoretical overpotentials ($\eta_{OER}$) for all the materials explored in this study as a function $\Delta G_{O^{*}} - \Delta G_{OH^{*}}$ and $\Delta G_{OH^{*}}$. The contour map is constructed using the scaling relation of $\Delta G_{OOH^{*}} = \Delta G_{OH^{*}} + 3.2$ eV. The green colored point is highlighted as the best material and facet unveiled by \textit{WhereWulff}.}
    \label{basrco_001_compare}
\end{figure}

Calculated surface energies, $\Delta G$ of \ce{OH^{*}}, \ce{O^{*}} and \ce{OOH^{*}}, the theoretical overpotential ($\eta_{oer}$) and the potential determining step (PDS) for all the considered surfaces and materials are summarized in Table~\ref{tab:dft_oer_reactivity}. As part of non-stoichiometric surface energy calculations, \ce{Ba} was consistently chosen as the reference species against which to compute free energy excess. Section S1 delves into the intricacies of this scheme. Following the work of Bajdich \textit{et al.}~\cite{cobalt_oxide_norskov}, the chemical potential for oxygen was obtained under alkaline conditions, with $\mu_{O} = G(\ce{H2O}) - G(\ce{H2}) + 2[eU - \Delta G_{H^{+}}(pH)] = -\text{5.02 eV}$, where $\Delta G_{H^{+}} = -k_{B} \cdot T \cdot \ln(10) \cdot pH$ and $T=$ 298.15 K. Given that \ce{Ba} species were taken into account as reference, it was required to compute the free energy excess for the \ce{Ti} and \ce{Co} species. In order to compute those, we optimized the bulk oxide of that transition metal species in which it exhibited the same oxidation state and coordination number as in the double perovskite, under a consistent level of theory. We then substracted off the energetics associated with the oxygen atoms using $\mu_{O}$ above, leaving us with $\mu_{Ti} = -\text{16.69 eV}$ and $\mu_{Co} = -\text{6.26 eV}$.
\input{tables/oer_dft}
Surface stability is also critical in determining the surface termination. The termination is likely to impact the local environment around the active site, resulting in changes in the adsorption interaction of the OER intermediates. As a result, it is important to establish the most stable surface coverage under reaction conditions to accurately describe the OER reactivity. Such surface coverage analysis is depicted for \ce{BaSrCo2O6}-(001) in Figure \ref{basrco_001_compare}-A. In this case, the clean termination is the most expressed termination in the applied potential range of 1.23 to 1.6 V$_{RHE}$.

We benchmarked our workflow on two well-studied materials: \ce{RuO2} and \ce{IrO2} (P4$_{2}$/mnm), which are the current state-of-the-art catalysts for OER in acidic conditions. Both the (110) and (101) are the most contributing crystallographic orientations to the nanoparticle shape, and hence are the focus of many theoretical studies\cite{gonzalez2021importance}. Surface Pourbaix diagrams for all these surface orientations describe that both \ce{RuO2} and \ce{IrO2} (110) and (101) surfaces are covered with \ce{O^{*}} under acidic OER conditions\cite{Acid-Stable-Norskov, gonzalez2021importance}. Thermodynamic OER overpotentials of 0.52/0.49 V and 0.53/0.39 V were obtained on top of \ce{O^{*}}-terminated \ce{RuO2}/\ce{IrO2} for the (110) and (101), respectively. For \ce{RuO2}/\ce{IrO2} (110) surfaces, the potential determining step in the associative OER mechanism is \ce{O^{*}} $\rightarrow$ \ce{OOH^{*}}, whereas for the (101) surface the potential determining step is \ce{OH^{*}} $\rightarrow$ \ce{O^{*}}. These results are in good agreement with values reported in the literature~\cite{Acid-Stable-Norskov,literature_kitchin,literature_montoya}.

Regarding the OER reactivity of the double perovskites, \ce{Ti} species were selected as the active site for \ce{BaSnTi2O6}, and \ce{Co} species for \ce{BaSrCo2O6} across the 3 most contributing crystallographic orientations. Corresponding atomistic visuals for \ce{BaSrCo2O6}-(001) OER intermediates, which is the closest benchmark to the results from Zheng \textit{et al.}~\cite{Hongliang} are depicted in Figure \ref{basrco_001_compare}-B. We refer the reader to the supporting information for more details on the rest of the materials and facets (see section S6 and S7). From the 2D OER activity volcano plot (Figure~\ref{basrco_001_compare}-C), we can see that \ce{BaSrCo2O6}-(001), \ce{BaSrCo2O6}-(101), \ce{BaSnTi2O6}-(001), \ce{BaSnTi2O6}-(110) and \ce{BaSnTi2O6}-(101) emerge as promising OER candidates that possess surfaces with relatively low theoretical overpotentials ($\eta_{OER}$). Among the promising OER materials, the \ce{BaSrCo2O6}-(001) has the lowest calculated overpotential of 0.63 V. For the \ce{BaSnTi2O6}-(101) surface, the potential determining step in the associative OER mechanism is \ce{O^{*}} $\rightarrow$ \ce{OOH^{*}} while the other two crystallographic orientations, (001) and (110), have their potential determining step as \ce{OH^{*}} $\rightarrow$ \ce{O^{*}}. On the other hand, for all the 3 selected \ce{BaSrCo2O6} surfaces, the potential determining step is \ce{OH^{*}} $\rightarrow$ \ce{O^{*}}, revealing the importance of the \ce{OH^{*}}. The superior performance of the \ce{BaSrCo2O6}-(001) facet can be attributed to the ease with which the \ce{OH^{*}} can be de-protonated~\cite{deprotonation}. The theoretical overpotential of \ce{BaSrCo2O6}-(001) agrees with previous literature~\cite{Hongliang}, which places it in the 0.5 - 0.75 V range. We also note that on the topic of the most stable coverage to perform reactivity on, for \ce{Ba5Sr5(Co6O17)2-(001)}, \ce{BaTi2SnO6}-(110) and \ce{Ba5Ti10Sn5O32-(101)}, we ran the reactivity under both terminations on either side of the dashed red reactivity line, acknowledging uncertainty in DFT calculations~\cite{venkat}. We saw that in the case of \ce{BaTi2SnO6}-(110), performing the activity analysis on top of a O$^*$-terminated surface yielded a theoretical overpotential of 2.06 V. However, for that same facet, when we performed the adsorption energy calculations on the OH$^*$-terminated surface, the theoretical overpotential improved to 0.73 V, highlighting the importance of capturing the local environment of the active site for that facet. On the other hand, testing both clean and OH$^*$-terminated surfaces for \ce{BaSrCo2O6}-(001) led to a negligible change from 0.63 to 0.62 V respectively. One explanation for the independence from surface coverage could have to do with the geometry of the (001) facet, where the Co sites are farther apart and the local interactions less significant than for the (110) facet.  We refer the reader to the SI for the rest of the visuals.
\onecolumn
\twocolumn \noindent

%% file: tables/oer_dft.tex
\begin{table*}
    \centering
    \resizebox{\textwidth}{!}{
    \begin{tabular}{ccccccccccccc}
    \hline
    \multirow{2}*{Formula}
              & (hkl) & $\gamma_{(hkl)}$ & Wulff & Coverage & $\Delta G_{OH*}$ & $\Delta G_{O*}$ & $\Delta G_{OOH*}$ & $\Delta G_{O*} - \Delta G_{OH*}$ & $\Delta G_{OOH*} - \Delta G_{O*}$ & $\eta_{oer}$ & Benchmark $\eta_{oer}$ & PDS  \\
              & & (J/m$^{2}$) & Contribution (\%) &  & (eV) & (eV) & (eV) & (eV) & (eV) & (V) & (V) & \\
              
    \hline
    \hline
    \multirow{2}*{\ce{RuO2}}
                & (110) & 0.96 & 49.2 & O* & 0.48 & 1.88 & 3.63 & 1.41 & 1.75 & 0.52 & 0.48~\cite{saha2020facet} & \ce{O^{*}} $\rightarrow$ \ce{OOH^{*}} \\
                & (101) & 1.05 & 50.8 & O* & 0.43 & 2.19 & 3.67 & 1.76 & 1.48 & 0.53 & 0.60~\cite{saha2020facet} & \ce{OH^{*}} $\rightarrow$ \ce{O^{*}} \\
    \hline
    \hline
    \multirow{2}*{\ce{IrO2}}
                & (110) & 1.33 & 46.3 & O* & 0.06 & 1.53 & 3.25 & 1.46 & 1.72 & 0.49 & 0.57\cite{Acid-Stable-Norskov} & \ce{O^{*}} $\rightarrow$ \ce{OOH^{*}} \\
                & (101) & 1.56 & 49.2 & O* & 0.59 & 2.21 & 3.71 & 1.62 & 1.50 & 0.39 & 0.41\cite{Acid-Stable-Norskov} & \ce{OH^{*}} $\rightarrow$ \ce{O^{*}} \\
    \hline
    \hline
    \multirow{3}*{\ce{BaSnTi2O6}}
                & (001) & 0.39 & 24.7 & OH* & 1.30 & 3.37 & 4.69 & 2.07 & 1.32 & 0.84 & NA & \ce{OH^{*}} $\rightarrow$ \ce{O^{*}} \\
                & (101) & 0.41 & 34.8 & OH* & 0.75 & 1.92 & 4.01 & 1.17 & 2.09 & 0.86 & NA & \ce{O^{*}} $\rightarrow$ \ce{OOH^{*}} \\
                & (110) & 0.66 & 30.9 & OH* & 1.47 & 3.43 & 4.67 & 1.96 & 1.24 & 0.73 & NA & \ce{OH^{*}} $\rightarrow$ \ce{O^{*}} \\
    \hline
    \hline
    \multirow{3}*{\ce{BaSrCo2O6}}
                & (001) & 0.33 & 27.2 & clean & 1.29 & 3.15 & 3.73 & 1.86 & 0.58 & 0.63 & 0.50 - 0.75~\cite{Hongliang} & \ce{OH^{*}} $\rightarrow$ \ce{O^{*}} \\
                & (101) & 0.38 & 15.7 & clean & 1.47 & 3.44 & 4.88 & 1.97 & 1.44 & 0.74 & NA & \ce{OH^{*}} $\rightarrow$ \ce{O^{*}} \\
                & (110) & 0.36 & 48.5 & clean & 1.26 & 3.49 & 4.73 & 2.23 & 1.24 & 1.00 & NA & \ce{OH^{*}} $\rightarrow$ \ce{O^{*}} \\
    \hline
    \end{tabular}}
    \caption{Results are shown for \ce{RuO2} and \ce{IrO2}, state-of-the-art OER catalysts whose primary purpose in the context of this paper is benchmarking. Having validated \textit{WhereWulff}, we summarize results for the unprobed materials (\ce{BaSrCo2O6} and \ce{BaSnTi2O6}) across all the prioritized facets that the workflow identified. Tabulated are the calculated surface energies (J/m$^{2}$), Wulff shape contribution (\%), binding free energies of \ce{OH^{*}}, \ce{O^{*}}, \ce{OOH^{*}} (eV), theoretical overpotential ($\eta_{oer}$) (V) and Potential Determining Step (PDS). A \ce{Ti} metallic center was selected to be the catalytic active site for \ce{BaSnTi2O6}, and \ce{Co} for \ce{BaSrCo2O6}.}
    \label{tab:dft_oer_reactivity}
\end{table*}

%% file: sections/conclusions.tex
\section{Conclusions}

In conclusion, we have shown that it is possible to encode a large portion of a traditional, rigorous and manual computational chemist workflow into a semi-autonomous workflow that handles the simulations, analyses and modeling for surface catalysis with a specific interface for OER. The workflow starts with a thorough optimization of a bulk material across dimensions of geometry and magnetic state followed by the bulk's stability characterization under reaction conditions.

That optimized bulk then feeds into our primary surface chemistry workflow called \textit{WhereWulff}, which performs an end-to-end reactivity analysis pipeline per material and per facet. This involves the creation of slab models and their prioritization based on Wulff nanoparticle shape contribution as a means of guiding their downstream adsorption energy tasks. This pipeline solves wall-time constraints that often plague these long running simulations in a way that ensures scale, provenance and efficiency.

To concretize these benefits, we apply the workflow to two previously unprobed perovskites, with the workflow being able to identify, with minimal human intervention, \ce{BaSrCo2O6}-(001) as the most promising material and facet. Based on the timing metadata collected by \textit{WhereWulff}, on average, we saved approximately 12 hours of re-submission time per material, per facet, assuming execution on CPU-only machines.

While previous work~\cite{Acid-Stable-Norskov} has enumerated the scientific steps required to rigorously model OER and there exists work~\cite{kravchenko2019new} of comparable magnitude focused on delivering efficient and didactic interfaces for catalytic studies, to the best of our knowledge, the formal encoding of OER domain expertise into an autonomous and scalable workflow and its release to the public was still a significant gap in the field up until this point.

Looking ahead, we highlight potential areas for future work, some of which are already under way. While \textit{WhereWulff} is autonomous and can technically scale to an arbitrary number of materials and facets, it still suffers from the computational cost of DFT. Substituting DFT with a surrogate model would get us closer to having a high-throughput workflow. This effort has already started with the recent release of pre-trained models based on OC22~\cite{OC22}, which we hope can yield raw energies and forces in a matter of seconds as opposed to days. We also hope that such surrogate models can allow us to implement more granular configurational searches for the lowest energy states than the ones that are currently performed in the workflow. Finally, we also acknowledge that surface stability is a function of the solvent environment: we plan on releasing a feature which allows users to see how the surface energy changes under various coverages and solvent saturation, based on ab initio thermodynamics, before performing the Wulff analysis.